\documentclass[a4paper,fleqn,usenatbib]{mnras}

\usepackage{mathptmx}

\usepackage[T1]{fontenc}
\usepackage{ae,aecompl}

\usepackage{graphicx}	
\usepackage{amsmath}	
\usepackage{amssymb}	
\usepackage{txfonts}
\usepackage{natbib}
\usepackage{lineno}
\usepackage{epsfig}
\usepackage{breqn}
\usepackage{fixltx2e}

\usepackage{bm}
\usepackage{siunitx}
\usepackage{array}
\usepackage{booktabs}
\usepackage{multirow}

\newcounter{address}
\newcommand{\degree}{\ensuremath{^\circ}}

\newcommand{\degsq}{\,deg$^{2}$}
\newcommand{\perdegsq}{\,deg$^{-2}$}
\newcommand\geqsim{\lower.73ex\hbox{$\sim$}\llap{\raise.4ex\hbox{$>$}}$\,$}
\newcommand\leqsim{\lower.73ex\hbox{$\sim$}\llap{\raise.4ex\hbox{$<$}}$\,$}
\newcommand{\project}[1]{\emph{#1}}

\newcommand{\sdssiv}{\project{SDSS-IV}}

\newcommand{\kms}{\,km\,s$^{-1}$}

\title[eBOSS ELG catalogs]{The \sdssiv~eBOSS: emission line galaxy catalogs at $z\approx0.8$ and study of systematic errors in the angular clustering}

\author[T. Delubac et al.]{
T. Delubac,$^{1}$\thanks{E-mail: timothee.delubac@epfl.ch}
A. Raichoor,$^{2}$
J. Comparat,$^{3,4}$
S. Jouvel,$^{5}$
J.-P. Kneib,$^{1,6}$
C. Y\`{e}che,$^{2}$
\newauthor
H. Zou,$^{7}$
J.R. Brownstein,$^{8}$
F.B. Abdalla,$^{5,9}$
K. Dawson,$^{8}$
E. Jullo,$^{6}$
A.D. Myers,$^{10}$
\newauthor
J.A. Newman,$^{11}$
W.J. Percival,$^{12}$
F. Prada,$^{13,14,4}$
A.J. Ross,$^{12,15}$
D.P. Schneider,$^{16,17}$
\newauthor
X. Zhou,$^{7}$
Z. Zhou,$^{7}$
and G. Zhu,$^{18}$
\\
$^{1}$Laboratoire d'Astrophysique, Ecole Polytechnique F\'{e}d\'{e}rale de Lausanne (EPFL), Observatoire de Sauverny, CH-1290 Versoix, Switzerland\\
$^{2}$CEA, Centre de Saclay, IRFU/SPP, F-91191 Gif-sur-Yvette, France\\
$^{3}$Instituto de F\'{\i}sica Te\'orica, (UAM/CSIC), Universidad Aut\'onoma de Madrid,  Cantoblanco, E-28049 Madrid, Spain\\
$^{4}$Departamento de Fisica Teorica, Universidad Autonoma de Madrid, Cantoblanco E-28049, Madrid, Spain\\
$^{5}$Department of Physics and Astronomy, University College London, Gower Street, London WC1E6BT, UK\\
$^{6}$Aix Marseille Univ, CNRS, LAM, Laboratoire d'Astrophysique de Marseille, Marseille, France\\
$^{7}$Key Laboratory of Optical Astronomy, National Astronomical Observatories, Chinese Academy of Sciences, Beijing, 100012, China\\
$^{8}$Department of Physics and Astronomy, University of Utah, Salt Lake City, UT 84112, USA\\
$^{9}$Department of Physics \& Electronics, Rhodes University, Grahamstown 6140, South Africa\\ 
$^{10}$Department of Physics and Astronomy, University of Wyoming, Laramie, WY 82071, USA\\
$^{11}$Department of Physics and Astronomy and PITT PACC, University of Pittsburgh, Pittsburgh, PA 15260, USA\\
$^{12}$Institute of Cosmology \& Gravitation, Dennis Sciama Building, University of Portsmouth, Portsmouth, PO1 3FX, UK\\
$^{13}$Institutode Astrof\'{i}sica de Andaluc\'{i}a (CSIC),Glorieta de la Astronom\'{i}a, E-18080 Granada, Spain\\
$^{14}$Campus of International Excellence UAM+CSIC, Cantoblanco, E-28049 Madrid, Spain\\
$^{15}$Center for Cosmology and Astroparticle Physics, Department of Physics, The Ohio State University, OH 43210, USA\\
$^{16}$Department of Astronomy and Astrophysics, The Pennsylvania State University, University Park, PA 16802\\
$^{17}$Institute for Gravitation and the Cosmos, The Pennsylvania State University, University Park, PA 16802\\
$^{18}$Department of Physics \& Astronomy, Johns Hopkins University, 3400 N. Charles Street, Baltimore, MD 21218, USA\\
}

\date{Accepted 2016 October 21. Received 2016 September 18; in original form 2015 December 21}

\pubyear{2016}

\begin{document}
\label{firstpage}
\pagerange{\pageref{firstpage}--\pageref{lastpage}}
\maketitle

\begin{abstract}
We present two wide-field catalogs of photometrically-selected emission line galaxies (ELGs) at $z\approx0.8$ covering about 2800\,\degsq~over the south galactic cap. 
The catalogs were obtained using a Fisher discriminant technique described in a companion paper. 
The two catalogs differ by the imaging used to define the Fisher discriminant: the first catalog includes imaging from the Sloan Digital Sky Survey and the Wide-Field Infrared Survey Explorer, the second also includes information from the South Galactic Cap $U$-band Sky Survey (SCUSS).
Containing respectively 560,045 and 615,601 objects, they represent the largest ELG catalogs available today and were designed for the ELG programme of the extended Baryon Oscillation Spectroscopic Survey (eBOSS).
We study potential sources of systematic variation in the angular distribution of the selected ELGs due to fluctuations of the observational parameters. We model the influence of the observational parameters using a multivariate regression and implement a weighting scheme that allows effective removal of all of the systematic errors induced by the observational parameters. We show that fluctuations in the imaging zero-points of the photometric bands have minor impact on the angular distribution of objects in our catalogs. We compute the angular clustering of both catalogs and show that our weighting procedure effectively removes spurious clustering on large scales. We fit a model to the small scale angular clustering, showing that the selections have similar biases of $1.35/D_a(z)$ and $1.28/D_a(z)$. Both catalogs are publicly available.
\end{abstract}




\section{Introduction}\label{sec:intro}

The development of large scale spectroscopic surveys such as the Sloan Digital Sky Survey~\citep[SDSS;][]{York00} and the 2-degree Field Galaxy Redshift Survey~\citep[2dFGRS;][]{Colless01} set a milestone in the era of precision cosmology by allowing the first detection of Baryon Acoustic Oscillations (BAO) in the power spectrum of the galaxy density field~\citep{Eisenstein05,Cole05}. The Baryon Oscillation Spectroscopic Survey~\citep[BOSS;][]{Dawson13}, one of the experiments of the third generation of the SDSS~\citep[SDSS-III;][]{Eisenstein11}, recently produced 1-2\% precision measurements of the BAO scale for redshifts $z<0.6$ using a total of about 1.3 million Luminous Red Galaxies (LRGs)~\citep{Tojeiro14,Anderson14}. It also produced a 2\% measurements of the BAO scale at redshift 2.34~\citep{Font14,Delubac15} using the Lyman-alpha forests of nearly 140,000 quasars. These results, combined with Cosmic Microwave Background~\citep{Planck14} and Supernova measurements~\citep[e.g.,][]{Betoule14}, yielded tight constraints on cosmological parameters such as the dark energy and dark matter densities, the dark energy equation of state, the Hubble parameter $H_0$ and the sum of neutrino masses $m_{\nu}$~\citep{Aubourg15}.

Building on the success of BOSS, the extended Baryon Oscillation Spectroscopic Survey~\citep[eBOSS;][]{Dawson16} will further tighten the current constraints on cosmological parameters by measuring the BAO scale at several redshifts in the currently unprobed range of $0.7<z<2.0$. This will be achieved by targeting LRGs up to redshift 1.0 but also by using two new tracers of the matter density field: quasar in the redshift range $0.9<z<2.0$ and Emission Line Galaxies (ELGs) in the redshift range $0.6<z<1.1$. eBOSS will provide the first 2\% measurement of the BAO scale using ELGs as a tracer~\citep{Zhao16}. This measurement will pave the way for future experiments such as the Dark Energy Spectroscopic Instrument\footnote{http://desi.lbl.gov/}~\citep[DESI;][]{Levi13}, the 4-metre Multi-Object Spectroscopic Telescope\footnote{https://www.4most.eu/}~\citep[4MOST;][]{deJong14}, the space mission Euclid\footnote{http://sci.esa.int/euclid/}~\citep{Laureijs11} of the European Space Agency (ESA) and the Subaru Prime Focus Spectrographe\footnote{http://sumire.ipmu.jp/en/}~\citep[PFS;][]{Takada14} that will all use ELGs as a main tracer of the matter density field. eBOSS uses the same facility as BOSS, including the 2.5\,m Sloan telescope~\citep{Gunn06}, the two fiber-fed spectrographs~\citep{Smee13} as well as an upgraded version of the BOSS pipeline~\citep{Bolton12}.

The development of large spectroscopic surveys is tightly coupled to the development of large photometric surveys such as the imaging programme of the SDSS~\citep{Gunn98,York00}, and the Canada France Hawaii Telescope Legacy Survey~\citep[CFHTLS;][]{Gwyn12}. Large and homogeneous photometric datasets are a requirement for the selection of spectroscopic targets. Since the early stages of the SDSS it has been recognised that inhomogeneities in the imaging datasets used for the target selection (e.g. due to variation of the seeing or sky flux) or astrophysical parameters such as the stellar density and Galactic extinction could induce artificial fluctuations in the target density and produce systematic errors in clustering measurements such as the angular or 3-dimensional correlation functions~\citep[see e.g.][]{Scranton02,Myers06,Ross11,Ho12}.

In this paper we present two large-scale catalogs of $z\approx0.8$ ELGs selected by their photometry using a Fisher discriminant technique~\citep{Fisher36}. The information used to produce the catalogs include SDSS $griz$ photometric bands, the Wide-field Infrared Survey Explorer\footnote{http://wise.ssl.berkeley.edu/} (WISE) $3.4\mu$m $W1$ band as well as the South Galactic Cap U-band Sky Survey\footnote{http://batc.bao.ac.cn/Uband/} (SCUSS) $U$ band. The selection algorithm as well as preliminary spectroscopic results obtained by eBOSS dedicated observations are presented in a companion paper~\citep[in the following R16]{Raichoor16}. The paper is structured as follow. Section~\ref{sec:requirements} lists the requirements to produce a catalog of ELGs suitable for a 2\% measurements of the BAO scale. Section~\ref{sec:imaging} presents the photometry used to produce the catalog and summarise the target selection algorithm and give information on how to access the catalogs. Section~\ref{sec:syst} describes a study of the impact of observational parameters (such as sky flux or seeing) and astrophysical parameters (such as stellar number densities or extinction) on the angular distribution of ELGs in our catalog. We use a multivariate regression to accurately model the fluctuations in the angular number density of ELGs induced by those parameters and propose a weighting procedure to remove them. Section~\ref{sec:clustering} presents the large scale angular clustering of our catalogs and show that our weighting procedure successfully remove spurious clustering on large scales. In section~\ref{sec:clusteringgal} we compute and model the small scale angular clustering of both catalogs and measure their biases. We conclude in Section~\ref{sec:conclusions}.

All magnitudes in this paper are given in the AB system~\citep{Oke83} and, unless stated otherwise, are corrected for Galactic extinction using dusts maps of~\citet{Schlegel98}. We consider a flat $\Lambda$CDM cosmology with $h=0.673$, $\Omega_mh^2 = 0.142$, $\Omega_bh^2 = 0.022$, $\sigma_8 = 0.829$, $n_s = 0.96$~\citep{Planck14} and a CMB temperature of $2.72$\,K. Coordinates are given in the J2000 equatorial system.

\section{Cosmological goals and implications for target selection}\label{sec:requirements}

The primary goal of the eBOSS ELG survey is to produce the first 2\% measurements of the BAO scale and redshift space distortion using ELGs as a tracer of the underlying matter density field. Constraints and requirements regarding the ELG survey are presented in eBOSS overview paper,~\citet{Dawson16}, while forecasts about cosmological results are given in~\citet{Zhao16}. Approximately 300 eBOSS plates will be dedicated to the ELG survey. The minimum surface density that can be efficiently tiled for the ELG programme is 170\,\perdegsq~thus the targets should be contained within a footprint of 1500\,\degsq. The final spectroscopic sample should contain classification of 190,000 ELGs, corresponding to a sample purity of 74\%. The width of the redshift distribution of the ELG sample should be less than $\Delta z = 0.4$ and contained in the redshift range $0.6<z<1.0$ or $0.7<z<1.1$. 

To ensure that our clustering analysis and cosmological measurements will be limited by statistical uncertainties, we further impose a list of requirements on our target sample that limit the systematic uncertainties. The requirements are derived from knowledge of the BOSS survey and are fully explained in~\citet{Dawson16}. We summarise the principal points below:
\begin{itemize}
\item The redshift uncertainty should be below 300\,\kms~RMS at all redshifts.
\item Catastrophic redshift errors (i.e. error exceeding 1000\,\kms and without knowledge that the redshift is wrongly estimated) should be below 1\%.
\item We require that the maximum absolute variation in expected galaxy density as a function of flux limit, stellar density, and Galactic extinction be less than 15\% within the survey footprint. 
\item We require that the maximum absolute variation in expected galaxy density as a function of imaging zero-point be less than 15\% within the survey footprint.  
\end{itemize}

\citet{Comparat16} studied in details the accuracy and catastrophic errors of ELG redshift assignment with eBOSS pipeline. The typical redshift accuracy is of the order of 30\,\kms, well below the requirement, and the catastrophic redshift fraction is lower than 1\%. R16 proposed two selections, both using a Fisher discriminant technique, that satisfy the target density, the redshift distribution, as well as the purity requirements. They developed the selection on a $\approx50$\,\degsq~area and evaluated it using dedicated eBOSS test plates. In this paper we investigate if these two selections, the UgrizW and griW selections, pass the remaining homogeneity requirements on large scales. In addition, we identify the dominant sources of systematic errors and investigate ways of correcting for them. 

\section{Parent imaging and target selection}\label{sec:imaging}

The UgrizW and griW selections both use combined information from different photometric catalogs. The griW selection uses SDSS $g$, $r$, and $i$ bands as well as WISE infrared $W1$ information, while the UgrizW also includes SDSS $z$-band as well as SCUSS $U$ band. This section introduces the different photometric catalogs used as well as the algorithms defining the selections. 

\subsection{SDSS imaging}

The SDSS is the principal imaging dataset used for our selections with all ELG targets being detected in SDSS images (while SCUSS and WISE counterparts are derived from forced-photometry). These images were obtained during the three first stages of the SDSS (I/II/III) using a wide-field imager~\citep{Gunn98} mounted on the SDSS 2.5 meters telescope~\citep{Gunn06}. Images were taken in the \emph{ugriz} system described in~\cite{Fukugita96}.
The 95\% detection repeatability for point sources of SDSS imaging has been estimated on the early data release of the SDSS to be on average $u=22.0$, $g=22.2$, $r=22.2$, $i=21.3$ and $z=20.5$~\citep{Stoughton02}. However those results are given in Luptitudes~\citep{Lupton99} and the south galactic cap (SGC) is not representative of the SDSS imaging as it is more contaminated by observational parameters such as extinction, which is on average higher than in the North Galactic Cap (NGC), and the airmass, which is higher for the SGC (at low declination). Accurate knowledge of the depth of the catalog is important as the targets are high-redshift galaxies and thus likely to reach the faint end of the catalog. 
To produce depth maps for each SDSS band we divide the SGC into equal area pixels using the HEALPix package\footnote{http://healpix.jpl.nasa.gov/}~\citep{Gorski99} with \texttt{NSIDE} set to 64 corresponding to pixels of 0.84\,\degsq. For each pixel, we select point-like objects (using SDSS flag \texttt{TYPE~=~6}) that have a signal-to-noise ratio for the \texttt{psfMag}\footnote{https://www.sdss3.org/dr10/algorithms/magnitudes.php\#mag\_psf} between 4 and 6. We then draw the histogram of the \texttt{psfMag} (uncorrected for extinction) of the objects in a given pixel and report the maximum of the histogram as the limiting magnitude. The resulting map for the $u$-band is shown on Fig.~\ref{fig:Udepth}. The 5\textsuperscript{th} percentile of the distributions of the limiting magnitude per pixel over the SGC footprint are, respectively, $u=21.6$, $g=22.8$, $r=22.3$, $i=21.8$ and $z=20.4$. In the following, unless otherwise stated, we consider \texttt{modelMag}\footnote{https://www.sdss3.org/dr10/algorithms/magnitudes.php\#mag\_model} magnitudes corrected for Galactic extinction for the SDSS photometric bands.

\subsection{SCUSS imaging}

\begin{figure*}
\begin{center}
\epsfig{figure=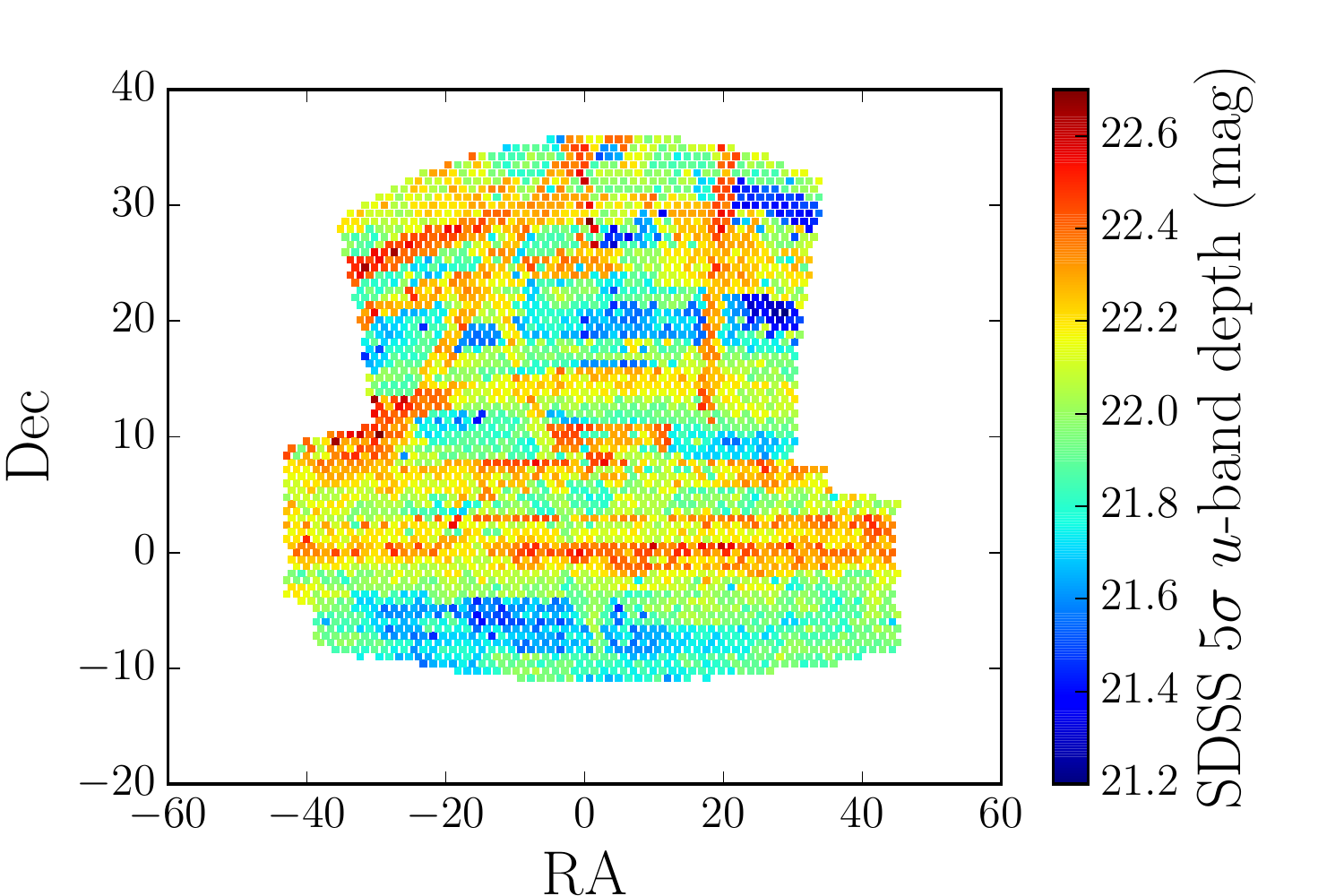,width=6cm}
\epsfig{figure=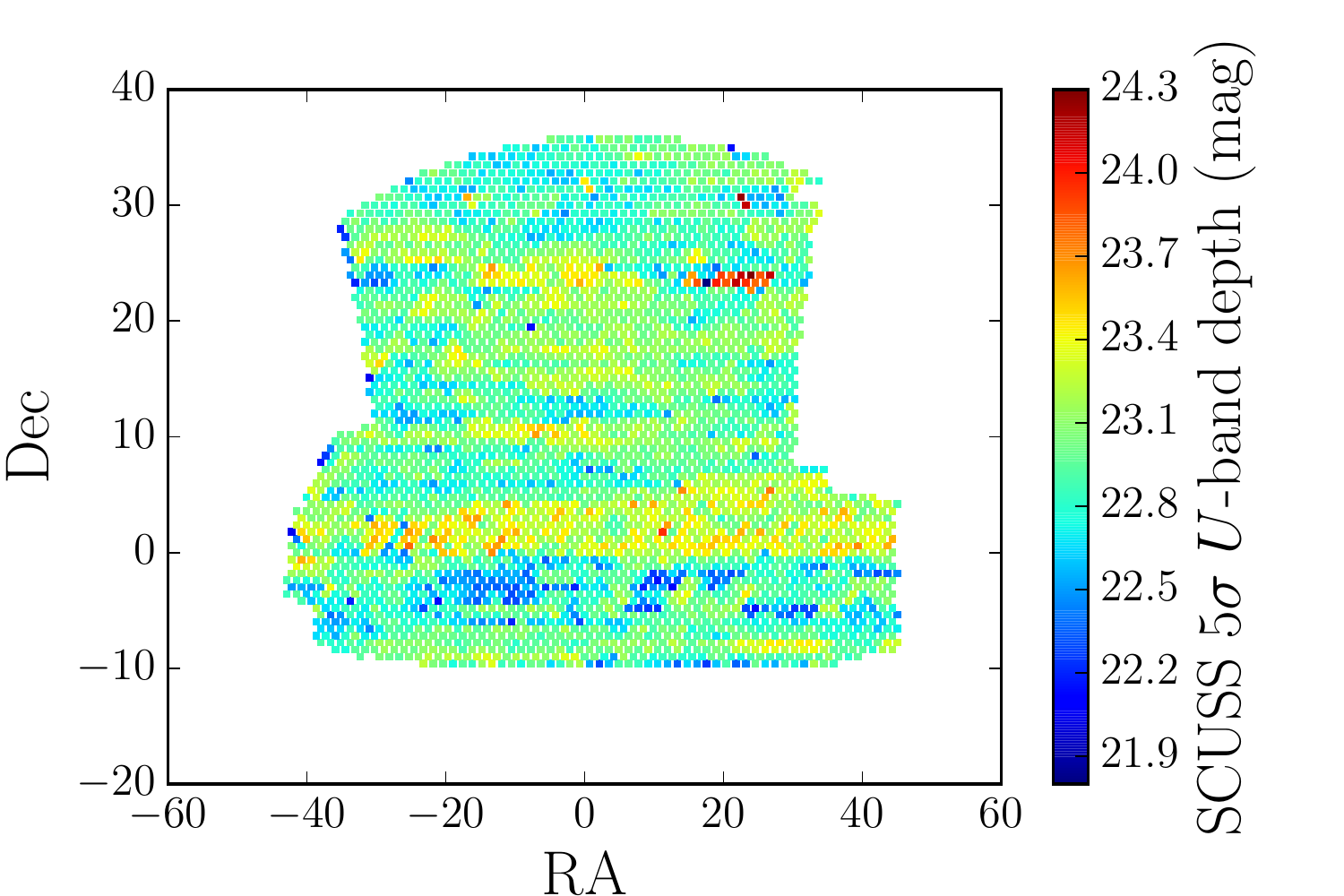,width=6cm}
\epsfig{figure=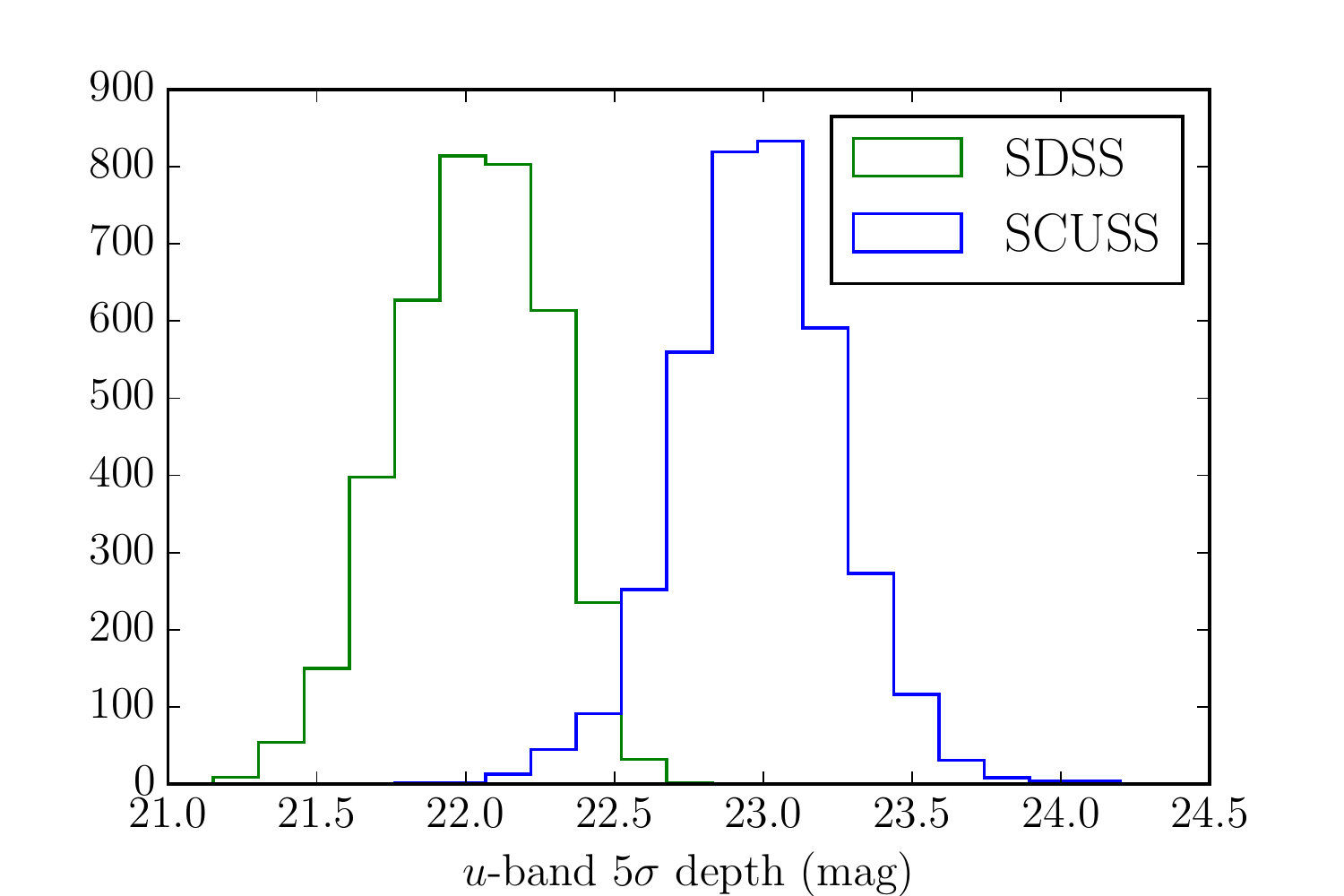,width=6cm}
\caption[]{\emph{Upper left:} 5$\sigma$ limiting magnitude maps of the SDSS SGC in the $u$-band. \emph{Upper right:} 5$\sigma$ limiting magnitude maps of SCUSS $U$-band. Both maps are divided into equal-area pixels of 0.84\,\degsq~(HEALPIX \texttt{NSIDE} $=$ 64). \emph{Bottom:} Histogram of the 5$\sigma$ limiting magnitude per pixel of SCUSS $U$-band (blue) and SDSS $u$-band (green). This histogram shows that the SCUSS survey is about one magnitude deeper than SDSS $u$-band.}
\label{fig:Udepth}
\end{center}
\end{figure*}

The South Galactic Cap $U$-band Sky Survey~\citep[SCUSS,][]{Zou15,Zou16} is an international project undertaken by the National Astronomical
Observatories of China (Chinese Academy of Sciences) and the Steward Observatory (University of  Arizona). It is a $U$-band ($\sim$3540$\,\AA$) survey that used the Bok 2.3\,m telescope to cover about 5000\,\degsq~of the SGC, 80\% of which overlaps the SDSS footprint. We use a forced-photometry catalog based on SDSS detections and, for each object, measure SCUSS \texttt{modelMag} using SDSS model to ensure consistant photometric measurements between SDSS and SCUSS. We estimate the SCUSS $5\sigma$ detection depth using the same technique as previously. The resulting map, together with the histogram of magnitude limit per pixel, is shown on Fig.~\ref{fig:Udepth}. Also presented is the histogram of the magnitude limit per pixel for SDSS $u$-band, which demonstrates that SCUSS is on average one magnitude deeper than SDSS $u$-band. This deeper $U$-band imaging is especially useful because the [\ion{O}{ii}] flux strength of ELGs correlates with the $u$-band magnitude~\citep{Comparat14}. 

\subsection{WISE imaging}

The Wide-Field Infrared Survey Explorer~\citep[WISE,][]{Wright10} is a full-sky survey in four mid-infrared wavelength filters labelled $W1$ to $W4$ centred at 3.4, 4.6, 12 and 22\,$\mu$m. We use the forced-photometry catalog of~\citet{Lang16}, where WISE photometry is derived using SDSS detections and galaxy profiles. As shown in R16, the use of WISE imaging allows efficient removal of low-redshift galaxies and can distinguish between $0.6\lesssim z_{\rm spec} \lesssim 1.0$ LRGs and ELGs. We only consider the $W1$ band as it has the highest signal-to-noise ration and as we found that including $W2$ band introduces spurious fluctuations in the density map likely due to Moon patterns in the $W2$ photometry. 

\subsection{Selection algorithm}

The method used to select ELGs is extensively described in the companion paper R16. It consists of a Fisher discriminant technique applied to a set of color measurements including SDSS $griz$-bands, SCUSS $U$-band and WISE $W1$-band information. The Fisher discriminant quantity $X_{FI}$ is defined as
\begin{dmath}\label{eq:Fisher}
X_{FI} = \alpha_0 + \alpha_{ur}\times(u-r)+\alpha_{gr}\times(g-r)+ \\ \alpha_{ri}\times(r-i) +\alpha_{rz}\times(r-z)+\alpha_{rW1}\times(r-W1),
\end{dmath}
where the values of $\alpha$ are optimised using training samples so that ELGs in the redshift range of interest receive a high value of $X_{FI}$ while other objects such as stars or galaxies outside the redshift range are assigned a small value of $X_{FI}$. Thus, a given Fisher selection is defined by a set of $\alpha$ parameters plus a lower cut on $X_{FI}$. R16 proposed two possible discriminants that satisfy eBOSS target selection requirements in terms of mean redshift, observation completeness, redshift accuracy and catastrophic redshift errors (those two last points being discussed in details in~\citealt{Comparat16}). The first selection is restricted to SDSS $gri$-bands and WISE $W1$-band, and is referred to as the griW selection. The second option includes information from the SDSS $z$-band and SCUSS $U$-band and is referred to as the UgrizW selection. We summarise the main characteristics of both selections in table~\ref{tab:Fisher}\footnote{The cut on the Fisher discriminant of the UgrizW selection is slightly lower than the one used in R16. This difference is due to the fact that the selection was optimised on a low extinction region on which the mean density of the selection is higher than on the full SGC. The cut was lowered to recover an average mean density greater than 180\,\perdegsq over the full SGC.} and refer the reader to R16 for a detailed discussion on the selections.

\begin{table*}
\begin{center}
\begin{tabular}{c|c|c}
& $griW$ & $UgrizW$\\
\hline
\hline
& & SCUSS $u$ \\
Photometric & SDSS $gri$ & SDSS $griz$\\
Datasets & WISE $W1$ & WISE $W1$  \\
&&\\

& $\alpha_0 = +0.104$ & $\alpha_0 = +0.956$ \\
& $\alpha_{ur} = 0$ & $\alpha_{ur} = -0.650$ \\
$X_{FI}$ param.& $\alpha_{gr} = -1.308$ & $\alpha_{gr} = -0.781$ \\
& $\alpha_{ri} = +0.870$ & $\alpha_{ri} = +0.065$ \\
& $\alpha_{rz} = 0$ & $\alpha_{rz} = +0.229$ \\
& $\alpha_{rW1} = +0.782$ & $\alpha_{rW1} = +0.739$ \\

&&\\

$X_{FI}$ cut & $X_{FI}>1.49$ & $X_{FI}>1.16$\\

&&\\

&&$20.0<u<23.5\mbox{, err}_u<1.0$  \\
&$20.0<g<22.5\mbox{, err}_g<0.5$ & $20.0<g<22.5\mbox{, err}_g<0.5$\\
Magnitude cuts&$19.0<r<22.5\mbox{, err}_r<0.5$ & $19.0<r<22.5\mbox{, err}_r<0.5$\\
&$19.0<i<21.5\mbox{, err}_i<0.5$ & $19.0<i<21.5\mbox{, err}_i<0.5$\\
&$17.0<W1<21.0\mbox{, err}_{W1}<0.5$ & $17.0<W1<21.0\mbox{, err}_{W1}<0.5$\\

&&\\

&\multicolumn{2}{c}{\texttt{BINNED2} = 0}\\
Other cuts &\multicolumn{2}{c}{\texttt{OBJC_TYPE} = 3 or $r>22$}\\
&\multicolumn{2}{c}{Bright stars masked}\\

\end{tabular}
\caption{The griW and UgrizW selection criteria for the catalogs in this paper. $\rm{err}_X$ corresponds to the estimated error on the magnitude $X$. For an extensive discussion on how those cuts were defined we refer the reader to R16. }
\label{tab:Fisher}
\end{center}
\end{table*}

\begin{figure*}
\begin{center}
\epsfig{figure=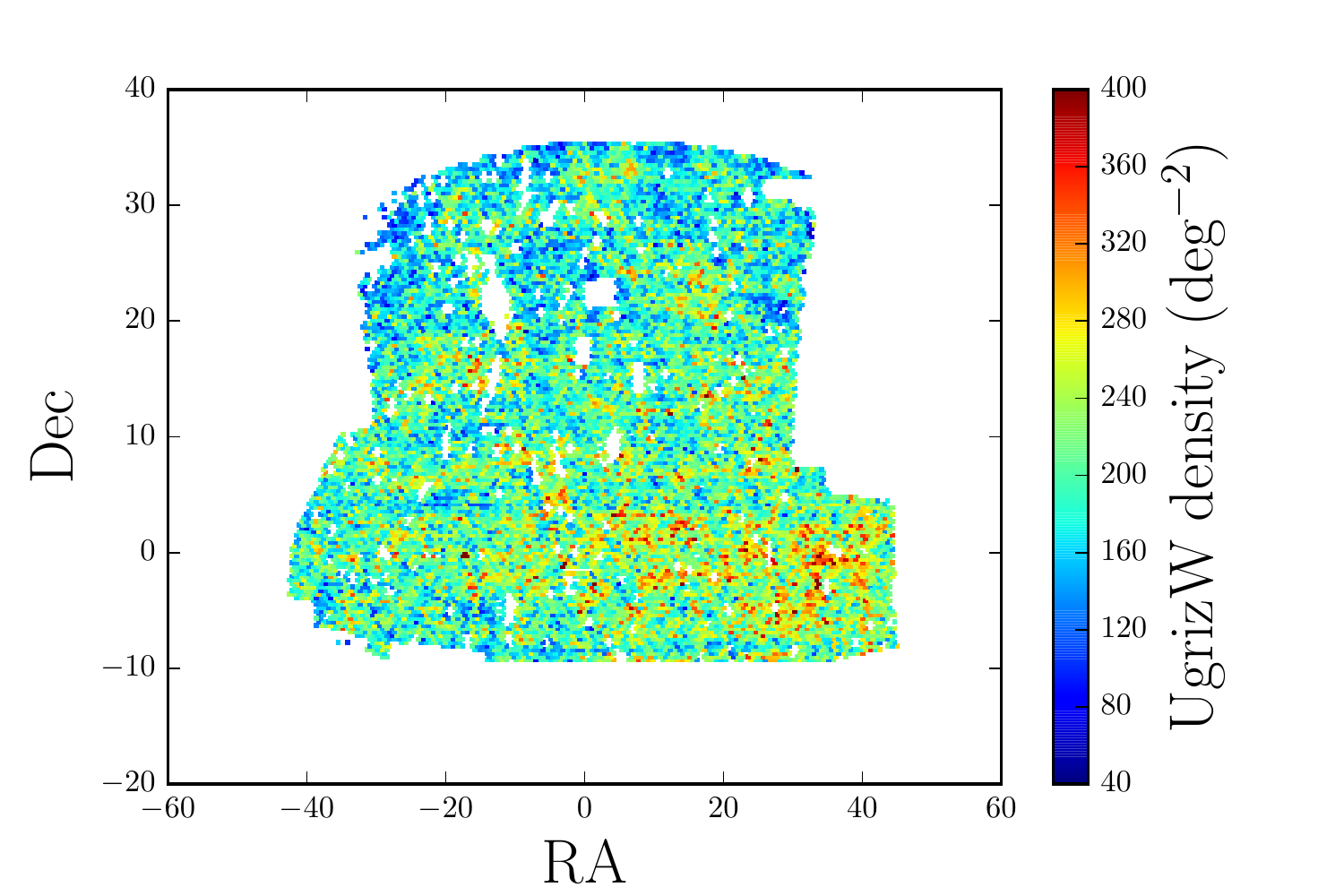,width=8cm}
\epsfig{figure=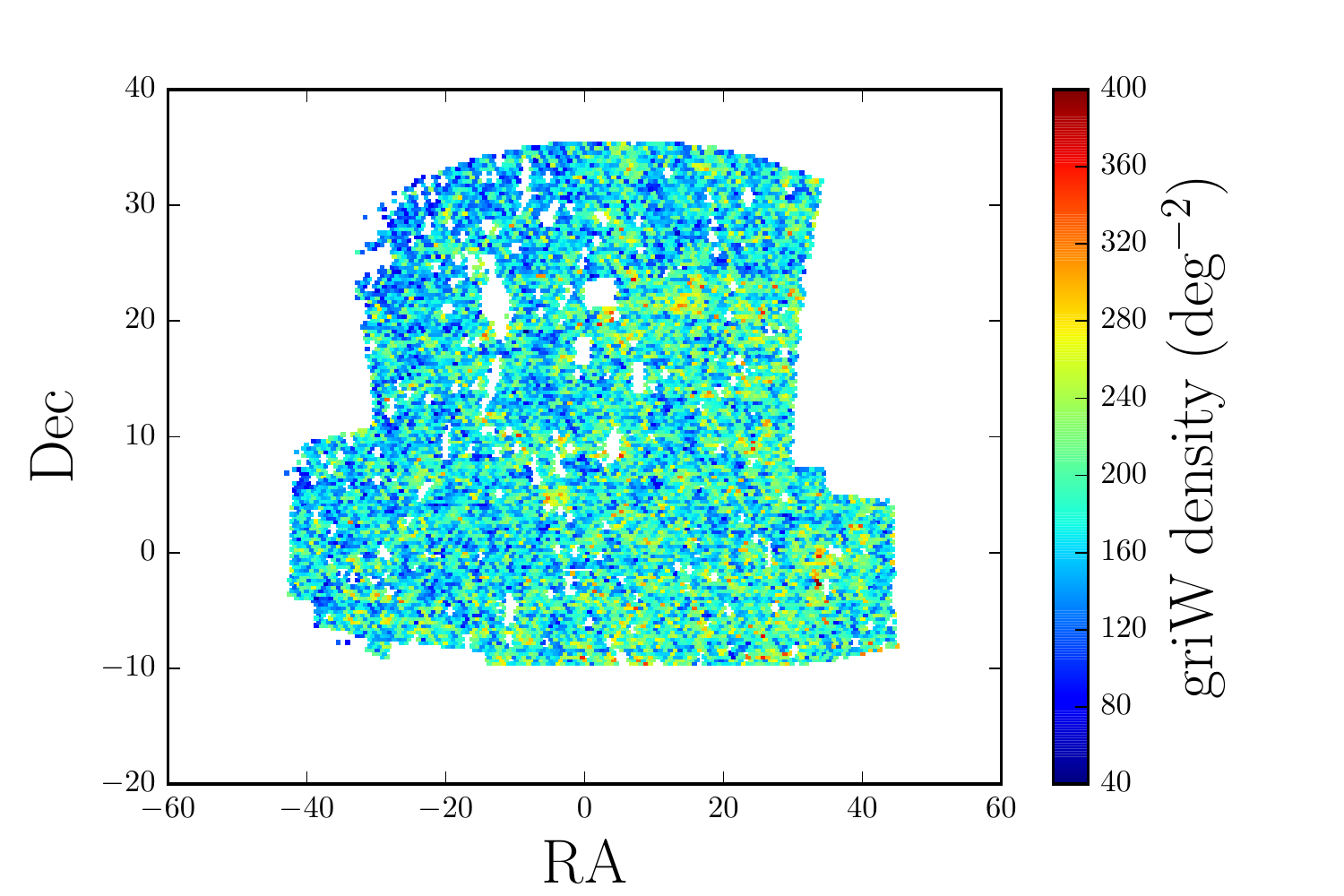,width=8cm}
\epsfig{figure=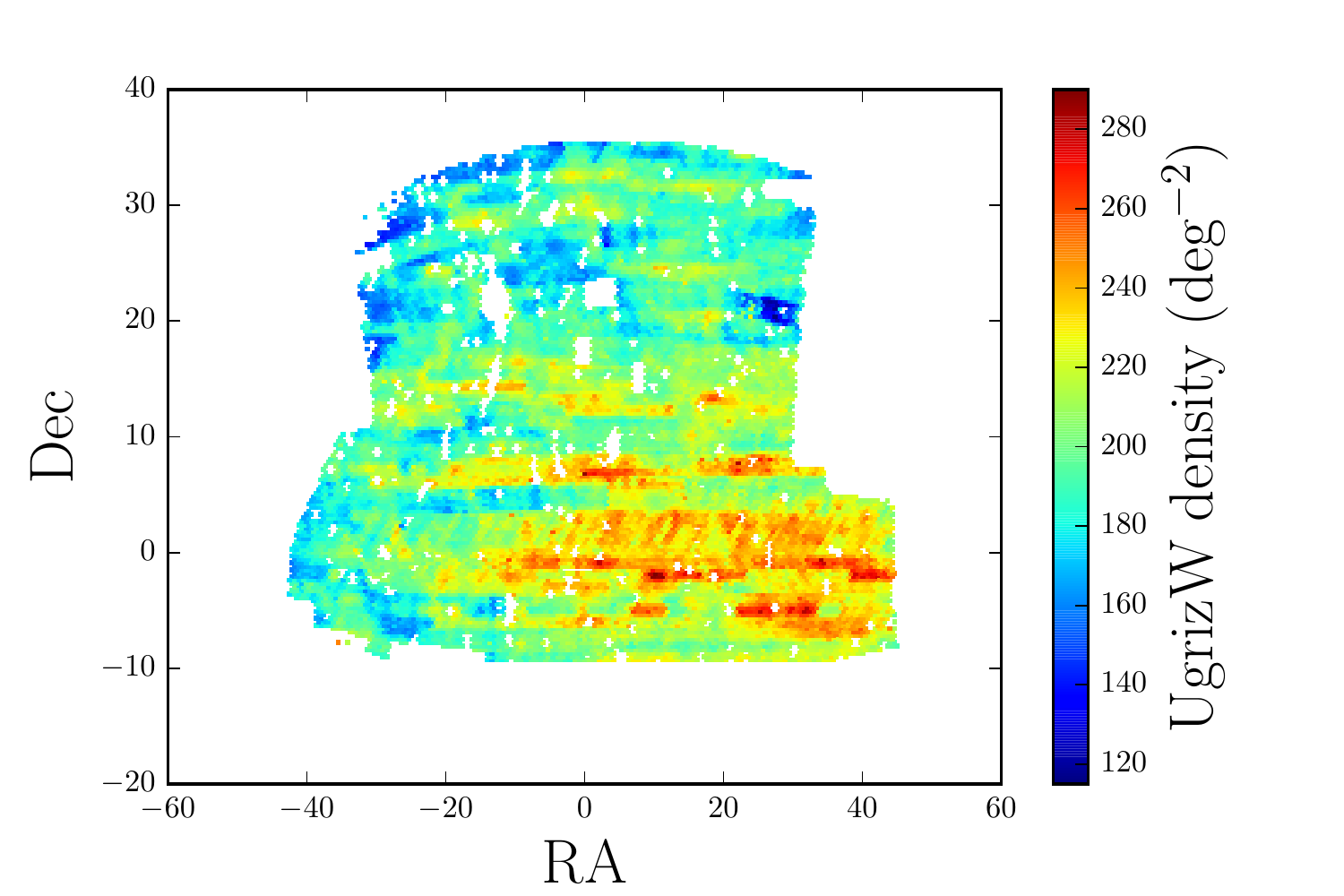,width=8cm}
\epsfig{figure=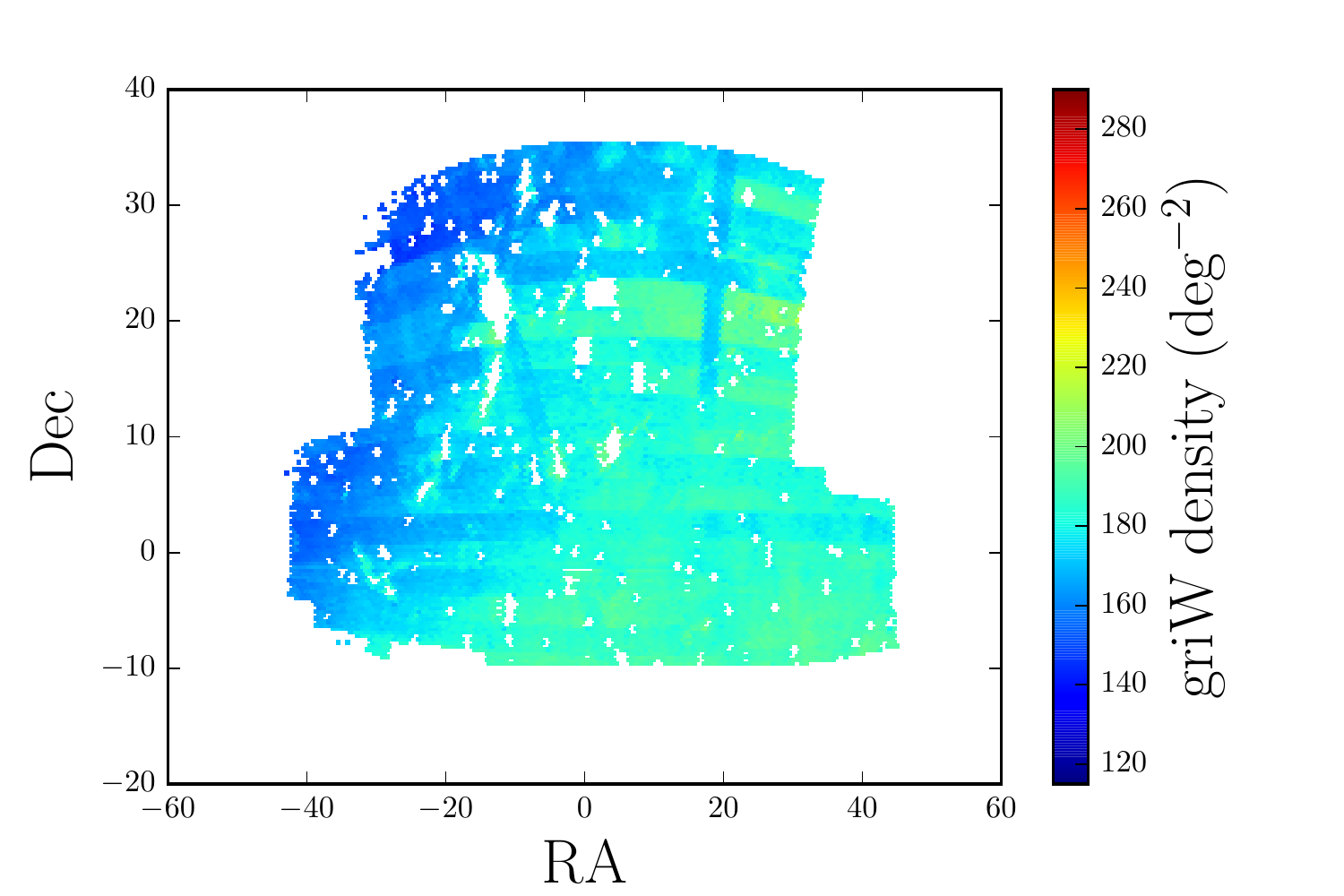,width=8cm}
\epsfig{figure=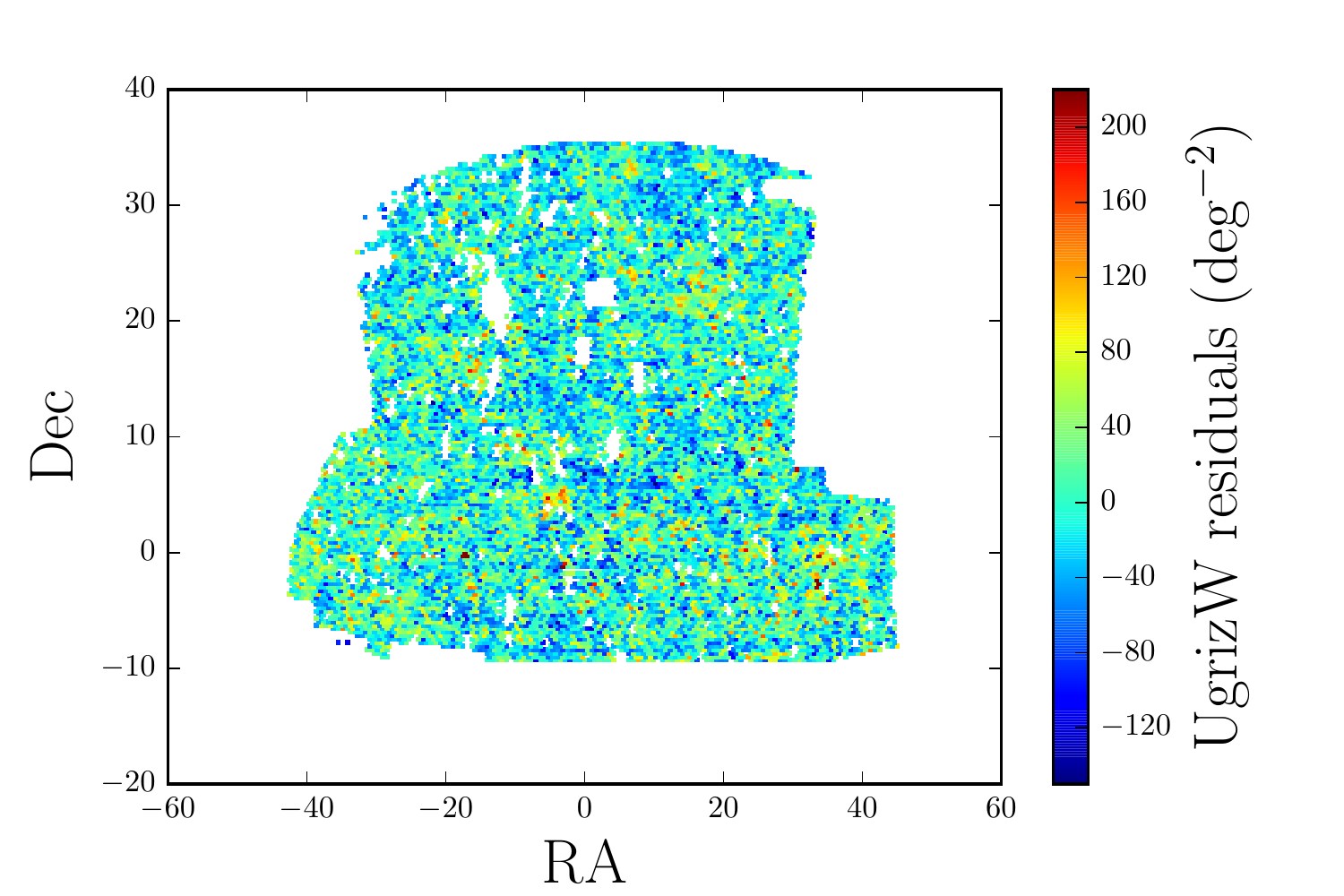,width=8cm}
\epsfig{figure=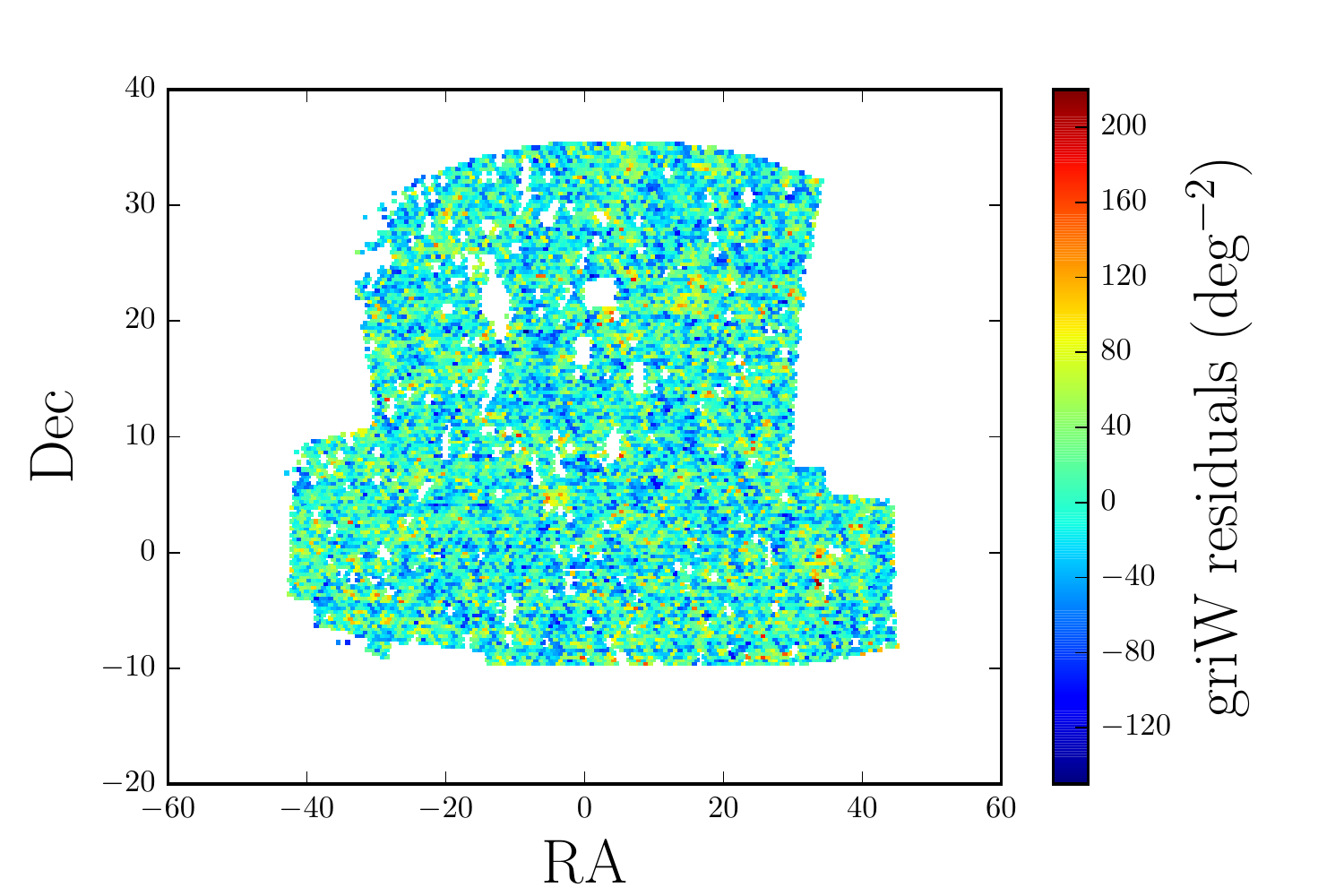,width=8cm}
\caption[]{\emph{Top row:} Density maps of the UgrizW (left) and griW (right) selections. The maps are divided in equal area pixels of 0.21\,deg$^{2}$ (HEALPix \texttt{NSIDE} = 128). The holes in the footprint correspond to masked regions. \emph{Middle row:} Predicted density maps for the UgrizW (left) and griW (right) selections obtained using the multivariate regression defined in Sec.~\ref{sec:modelsyst}. The good agreement with the original density maps is clearly visible as higher density regions in the selection density maps correspond to high density regions in the predicted density maps. We emphasize that the color scale of the griW and UgrizW maps are the same, clearly demonstrating that the griW predicted density is more homogeneous than the UgrizW one. \emph{Bottom row:} Residual maps for the UgrizW (left) and griW (right) selections obtained by subtracting the predicted densities to the selection densities.}
\label{fig:densmaps}
\label{fig:predicteddens}
\label{fig:residuals}
\end{center}
\end{figure*}

\subsection{Accessing the catalog} 

The UgrizW and griW catalogs are publicly available\footnote{https://data.sdss.org/sas/dr13/eboss/target/elg/fisher-selection/}. The catalogs include for each object position and photometric information as well as estimates of the correction weight defined in Sec.~\ref{sec:weights} for clustering analysis. Though we do not study systematics on the North Galactic Cap (NGC) in this paper, the public version of the griW selection also includes objects selected on the NGC, corresponding to 1,440,750 more objects. The UgrizW catalog does not contain object over the NGC as no SCUSS information is available on that region. A full description of the catalogs is available on the SDSS website\footnote{https://data.sdss.org/datamodel/files/EBOSS_TARGET/elg/fisher-selection/}. 

\section{Systematic effects on the angular distribution of galaxies}\label{sec:syst}

\subsection{Observational parameters}

The griW and UgrizW catalogs are cleaned using standard SDSS Mangle~\citep{Swanson08} masks for bright stars, bright objects and bad photometry (\texttt{bright_stars}, \texttt{badfield}, \texttt{bad_objects})\footnote{http://data.sdss3.org/sas/dr10/boss/lss/reject\_mask/}, as well as an additional mask intended to remove bright stars surroundings in WISE photometry.  The latter is a custom mask that was developed after finding over-densities in our selection in the vicinity of bright stars. These over-densities resulted from the over-estimation of WISE $W1$ fluxes in the halo of bright stars on radii significantly larger than the one removed by SDSS masks. In its current implementation, described in R16, this custom mask removes about 100\,\degsq~over the SGC. This estimate is conservative, and in the future a more careful estimation of the span of $W1$ bright star halos could reduce the masked area. 

After applying masks, we divide our maps in equal area pixels of $\approx 0.21$\,\degsq~(corresponding to HEALPix \texttt{NSIDE = 128}), and compute the number density in each pixel. The resulting maps are shown in the top row of~Fig.~\ref{fig:densmaps}. The choice of the size of the pixels is a compromise between the necessity of having sufficiently small pixels to have a precise estimation of the observational parameters, and the requirement of having sufficiently large pixels such that we do not have to deal with empty pixels and that the shot noise can be approximated as Gaussian. We also require that the pixel size be smaller than the BAO scale of 2.99$\degree$ in our fiducial cosmology for $z=0.8$.  We compute the effective area of each pixel by accounting for area covered by the different masks. We remove every pixel that has more than 20\% of its area masked and otherwise correct the density by the amount of unmasked area. 

For both the griW and UgrizW selections, we consider the following observational parameters as potentially impacting the angular distribution of galaxies: 
\begin{itemize}
\item $n_{stars}$: The number density of SDSS \texttt{TYPE = 6} objects (point sources) satisfying $18<g<21$ where $g$ corresponds to a point spread function magnitude not corrected for Galactic extinction. We select stars in the $g$-band to be consistant with our $g$-band limited selections, but without going as deep as 22.5 as morphological informations are not reliable at that level. We also cut on the \texttt{CLEAN} flag\footnote{https://www.sdss3.org/dr8/algorithms/photo\_flags\_recommend.php} to include only secure photometric detections.
\item $A_g$: $g$-band extinction derived from Schlegel et al. (1998) Galactic extinction maps and converted into magnitudes using the relation $A_g = 3.303E(B-V)$~\citep{Dawson16}.
\item airmass\textsubscript{SDSS}: SDSS airmass in $g$-band.
\item sky\textsubscript{SDSS}: SDSS sky flux\footnote{https://www.sdss3.org/dr10/algorithms/magnitudes.php\#nmgy} in $g$-band in nanomaggies.arcsec$^{-2}$.
\item seeing\textsubscript{SDSS}: SDSS FWHM seeing in arcsec. 
\item covmedian: WISE median number of single-exposure frames per pixel.
\end{itemize}
For the UgrizW selection, we consider the following additional parameters:
\begin{itemize}
\item sky\textsubscript{SCUSS}: SCUSS sky flux in magnitude. 
\item seeing\textsubscript{SCUSS}: SCUSS seeing in arsec.
\end{itemize}
All of those observational parameters are summarised in tab.~\ref{tab:syst}.

\begin{table*}
\begin{center}
\begin{tabular}{
	c
	c
	c
	c
	S[table-format=3.2]
	S[table-format=3.2]
	c}
\toprule
Obs. parameter & Survey & unit & cut & {$\bar{s}_i$} & {$\sigma_{s_i}$} & model\\
\midrule
$n_{stars}$ & all & deg$^{-2}$ & $<$ 6000 & {1.9e3} & {1.1e3} & Q\\
$A_g$ & all & mag. & $<$ 0.5 & 0.19 & 0.085 & Q\\
sky\textsubscript{SDSS} & SDSS & nanomaggies.arcsec$^{-2}$ &  $<$ 2.4 & 1.6 & 0.25 & Q\\
airmass\textsubscript{SDSS} & SDSS & - & - & 1.3 & 0.19 & L \\
seeing\textsubscript{SDSS} & SDSS & arcsec. & - & 1.3 & 0.18 & L\\
covmedian & WISE & - & - & 28 & 5.4 & L\\
seeing\textsubscript{SCUSS}\textdagger& SCUSS & mag. & - & 1.9 & 0.40 & L\\
sky\textsubscript{SCUSS}\textdagger & SCUSS & mag. & - & 22 & 2.2 & L\\
 \bottomrule
\end{tabular}
\caption{Observational parameters studied for the two selections. The \emph{Survey} column indicates to what survey the parameter is relevant. The \emph{cut} column provides, when relevant, the cuts applied for the analysis. The $\bar{s}_i$ and $\sigma_{s_i}$ columns, respectively, list the mean and the standard deviation of the value of the observational parameters in the analysis pixels. The \emph{model} column indicates whether the modelling of the parameter in Eq.~\ref{eq:syst_model} is quadratic (Q) or linear (L). Parameters marked with a \textdagger~symbol are only considered for the UgrizW selection.}
\label{tab:syst}
\end{center}
\end{table*}

\begin{figure*}
\begin{center}
\epsfig{figure=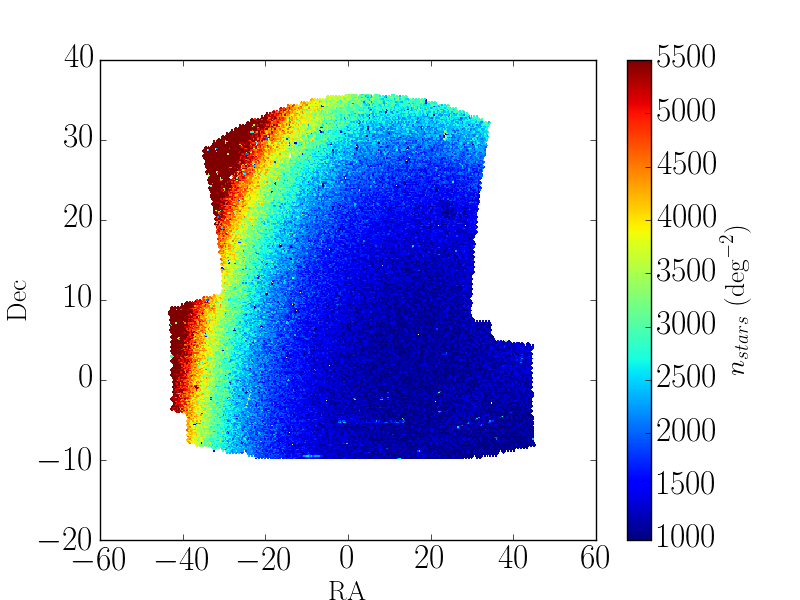,width=7.3cm}
\epsfig{figure=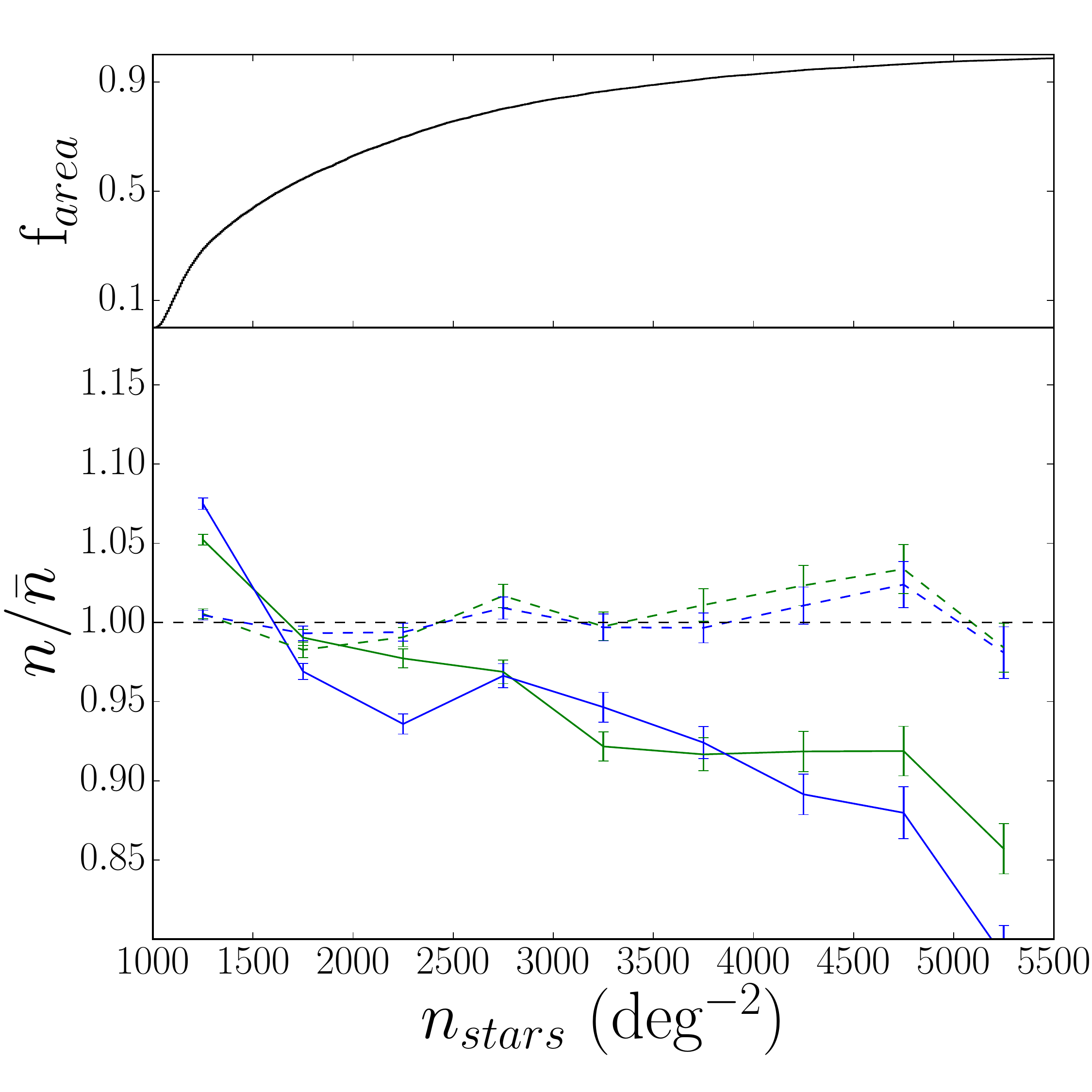,width=5.3cm}

\epsfig{figure=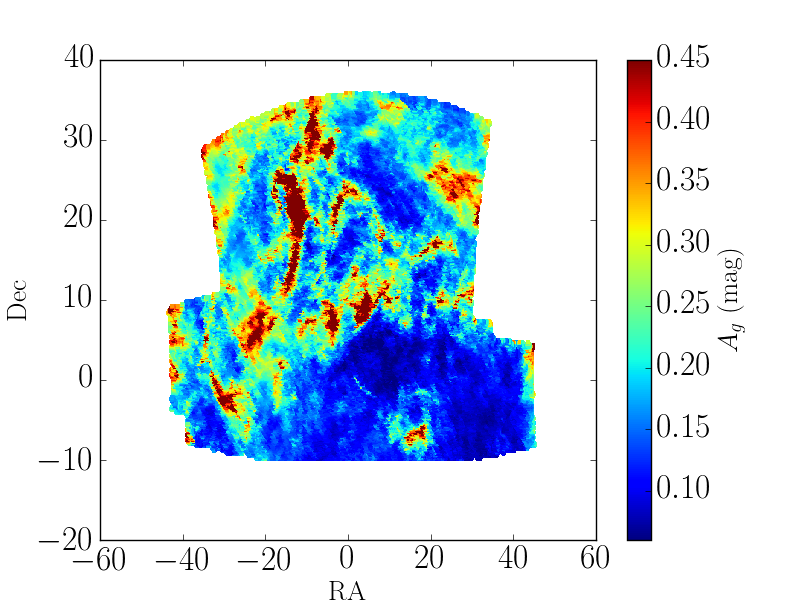,width=7.3cm}
\epsfig{figure=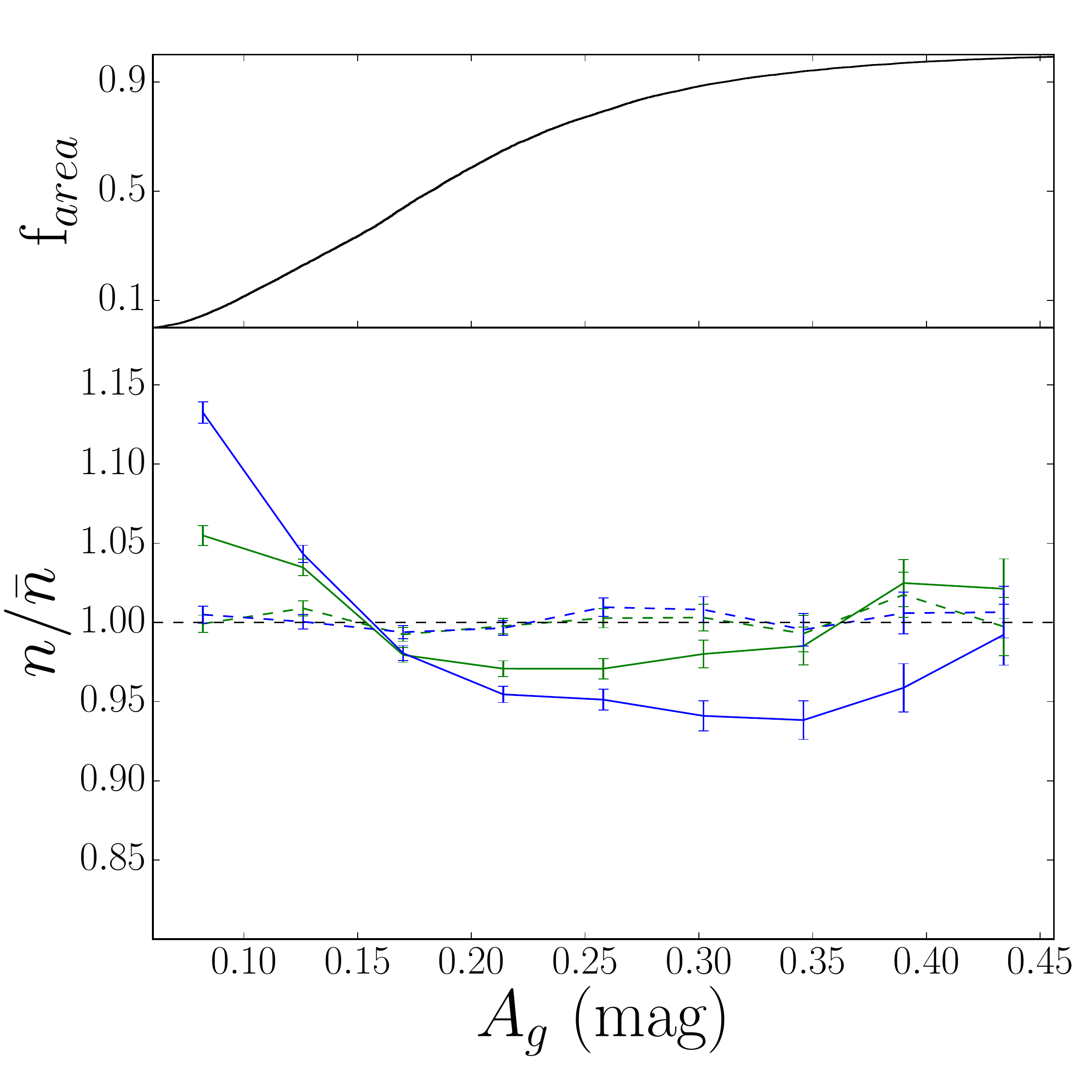,width=5.3cm}

\epsfig{figure=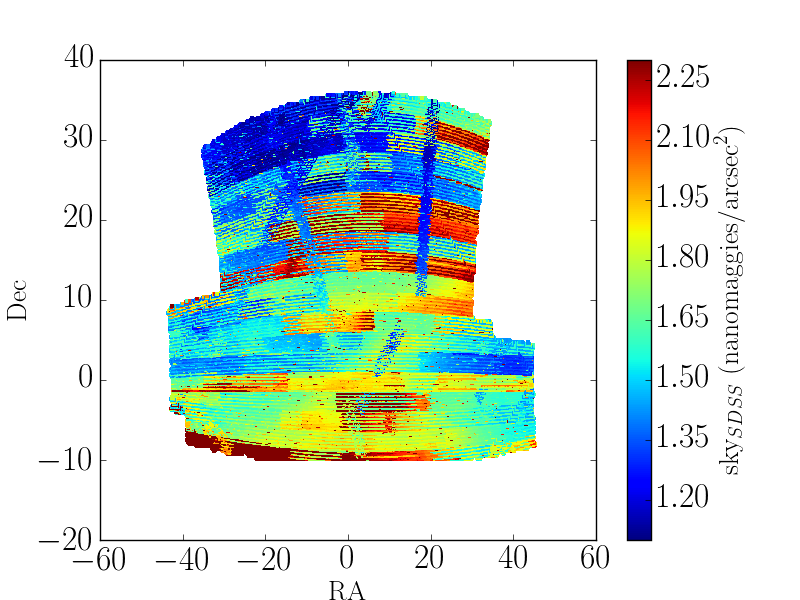,width=7.3cm}
\epsfig{figure=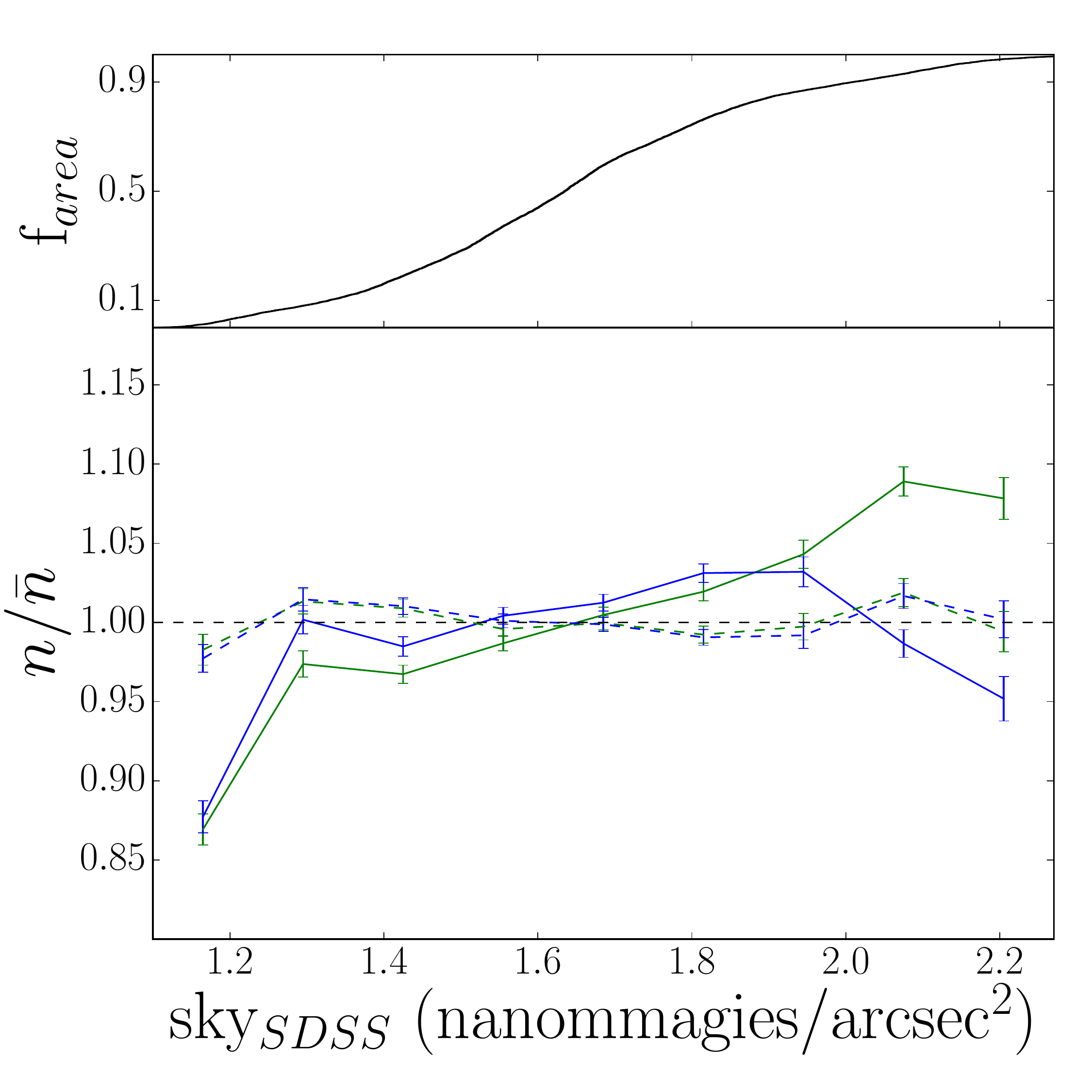,width=5.3cm}

\epsfig{figure=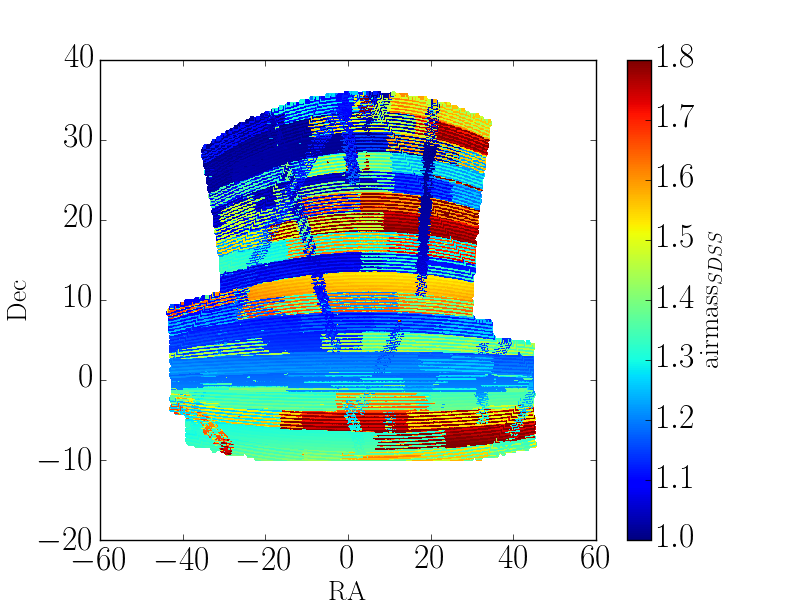,width=7.3cm}
\epsfig{figure=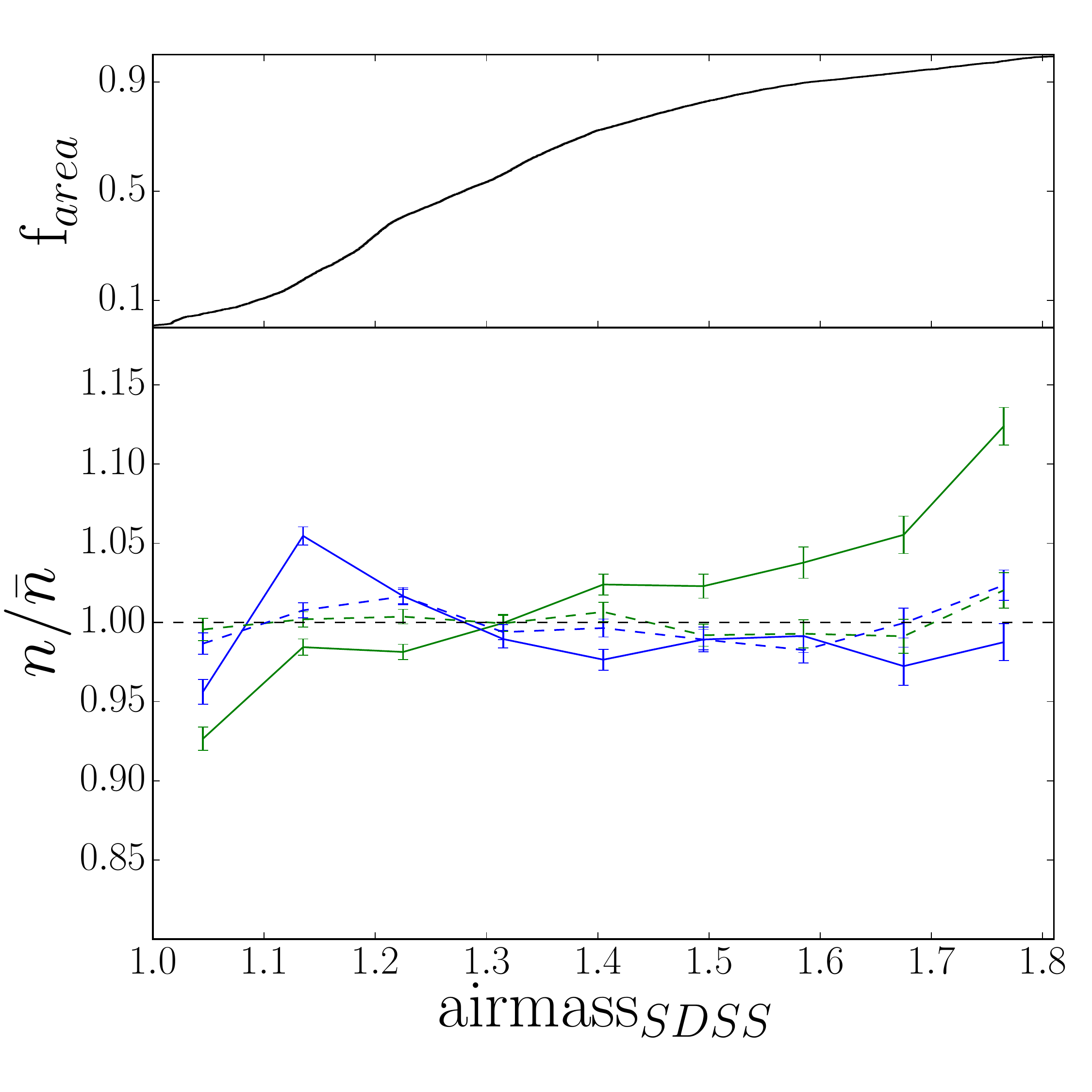,width=5.3cm}

\caption[]{Maps of the observational parameters (left) and evolution of the normalised average number density ($n/\bar{n}$) as a function of each observational parameter (right). On each panel on the right, the top curve shows the fractional area of the survey that has a value of the parameter lower or equal to the $x$-axis value. On the bottom panels solid lines correspond to the uncorrected density fluctuations, while dashed curves represent the fluctuations remaining after applying the weights defined in Sec.~\ref{sec:weights}. Blue curves are for the UgrizW selection and green curves for the griW selection. From top to bottom the observational parameters correspond to the stellar density of stars with $18<g<21$, the Galactic extinction in $g$-band ($A_g$), SDSS sky flux (sky\textsubscript{SDSS}) and SDSS airmass (airmass\textsubscript{SDSS}). 
}
\label{fig:systmaps}
\end{center}
\end{figure*}

\begin{figure*}
\begin{center}
\epsfig{figure=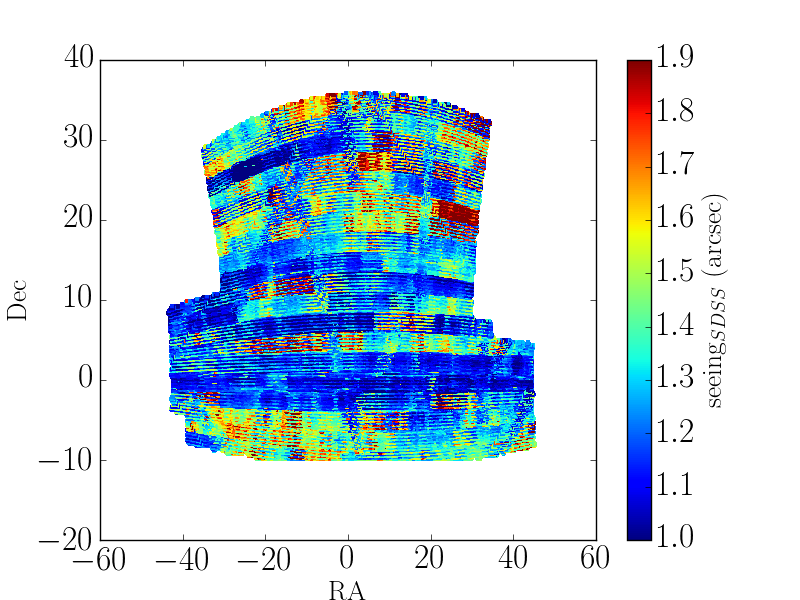,width=7.3cm}
\epsfig{figure=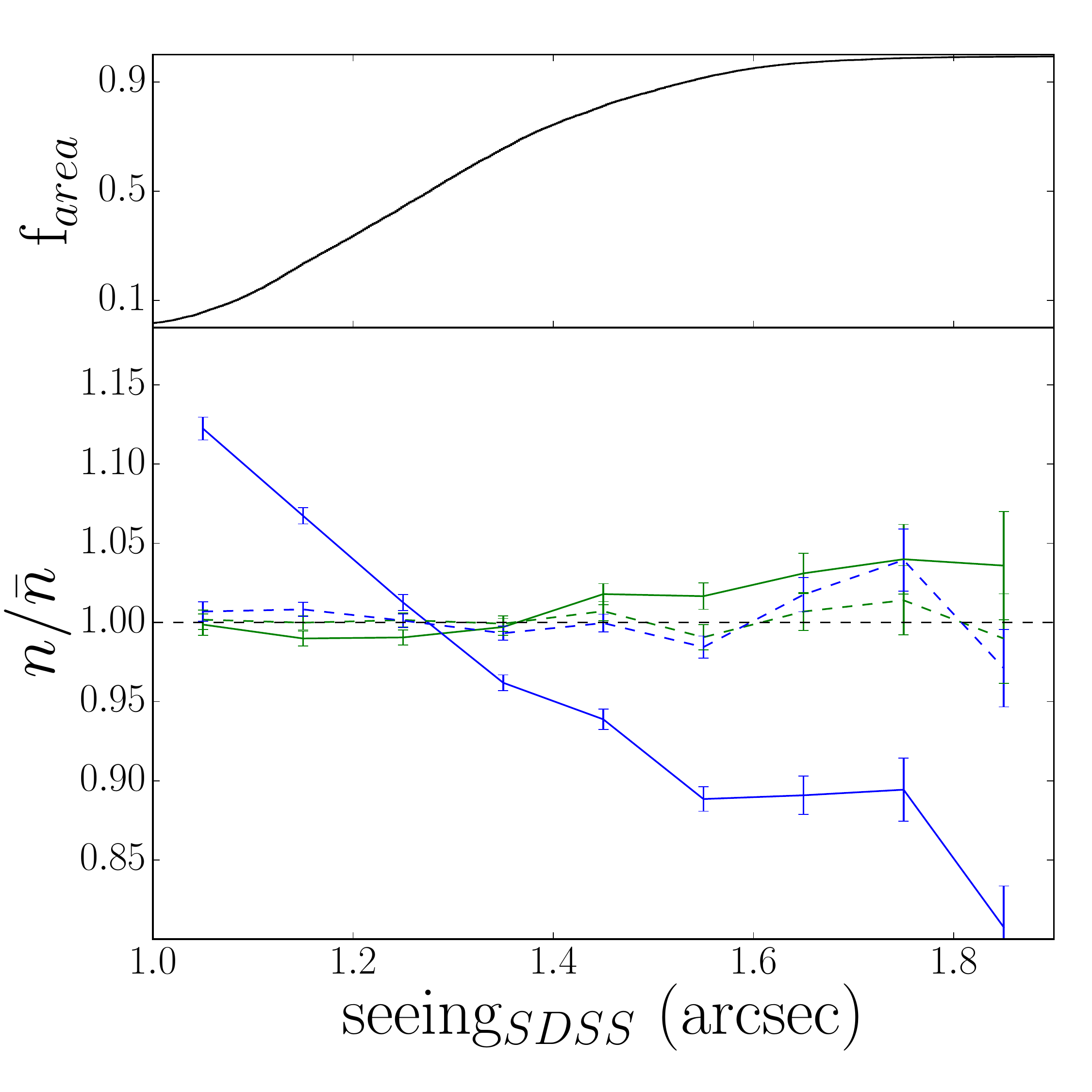,width=5.3cm}

\epsfig{figure=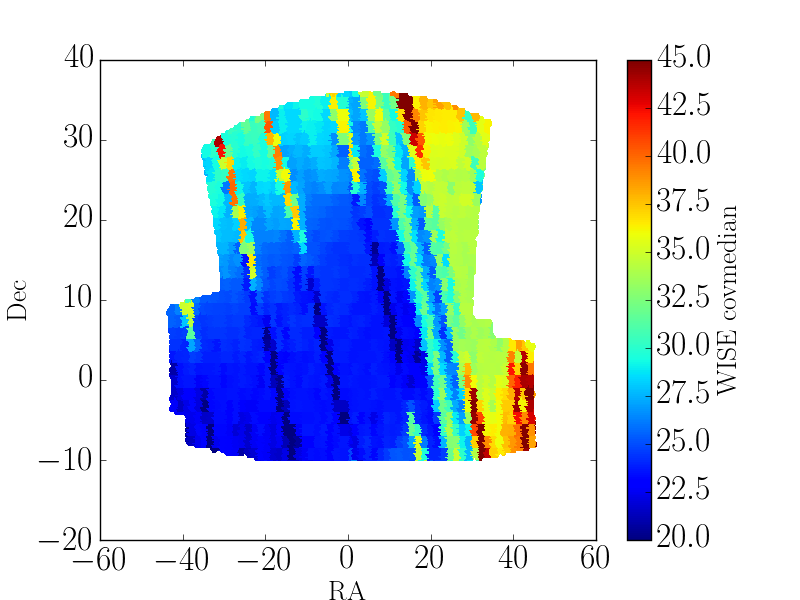,width=7.3cm}
\epsfig{figure=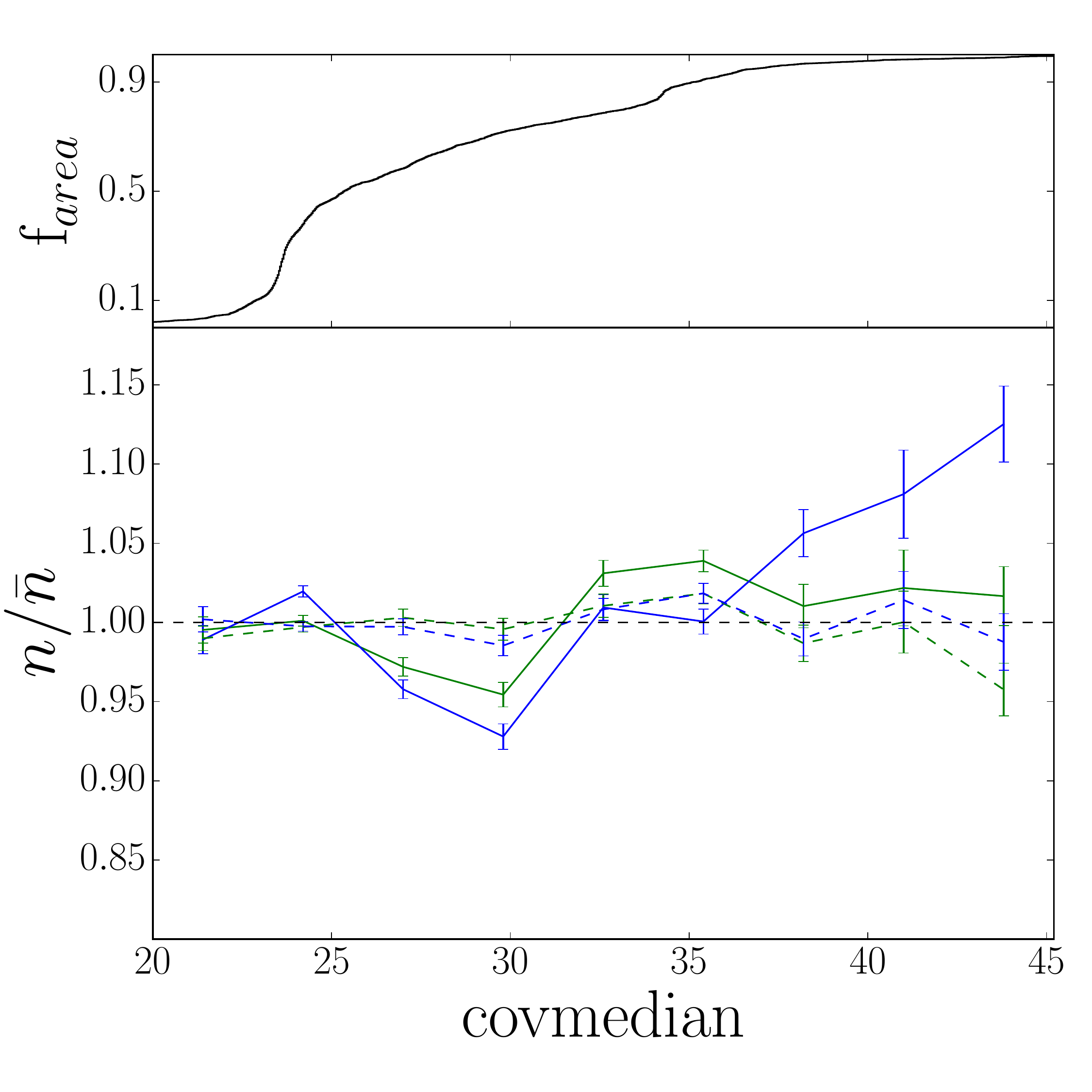,width=5.3cm}

\epsfig{figure=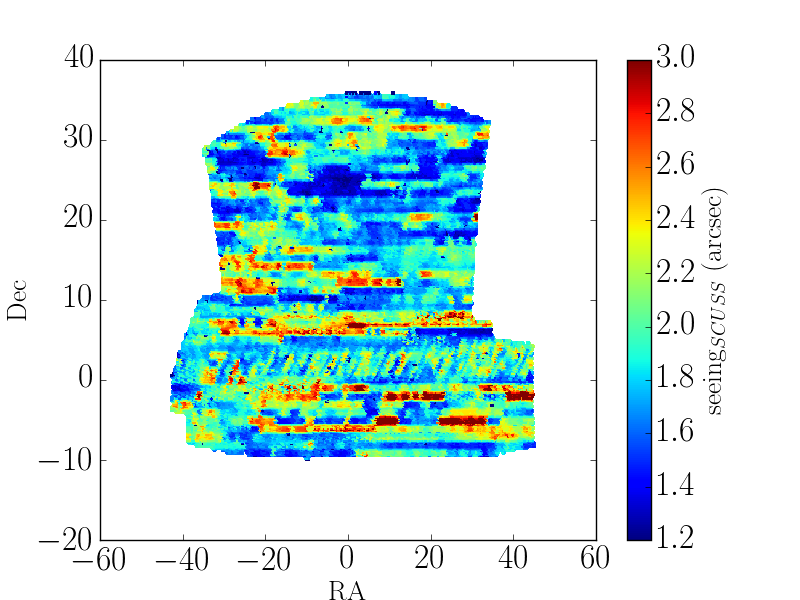,width=7.3cm}
\epsfig{figure=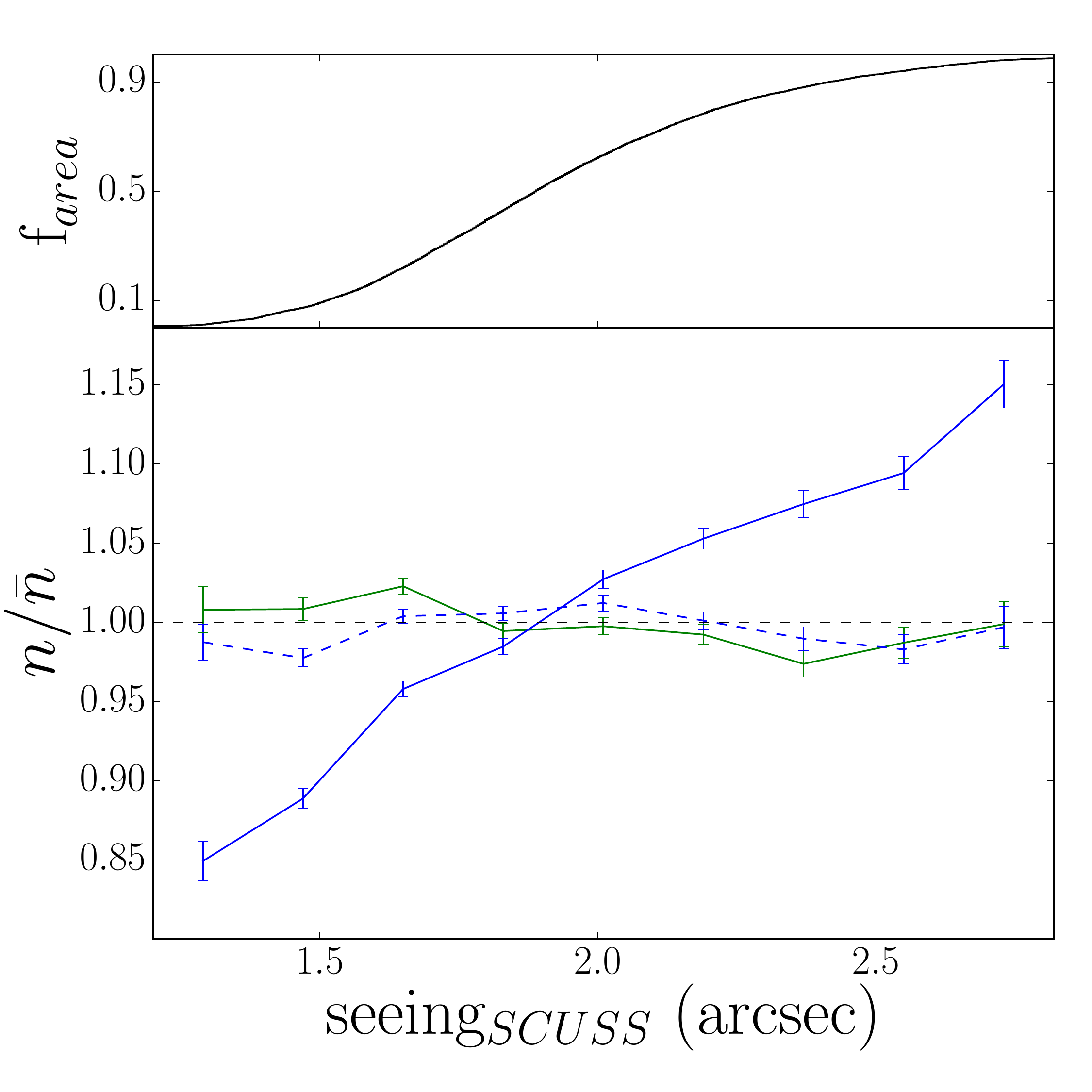,width=5.3cm}

\epsfig{figure=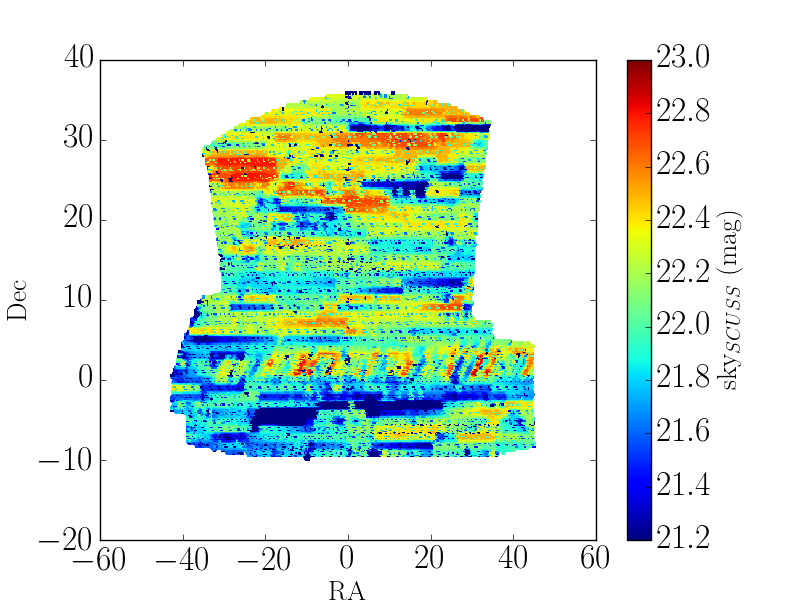,width=7.3cm}
\epsfig{figure=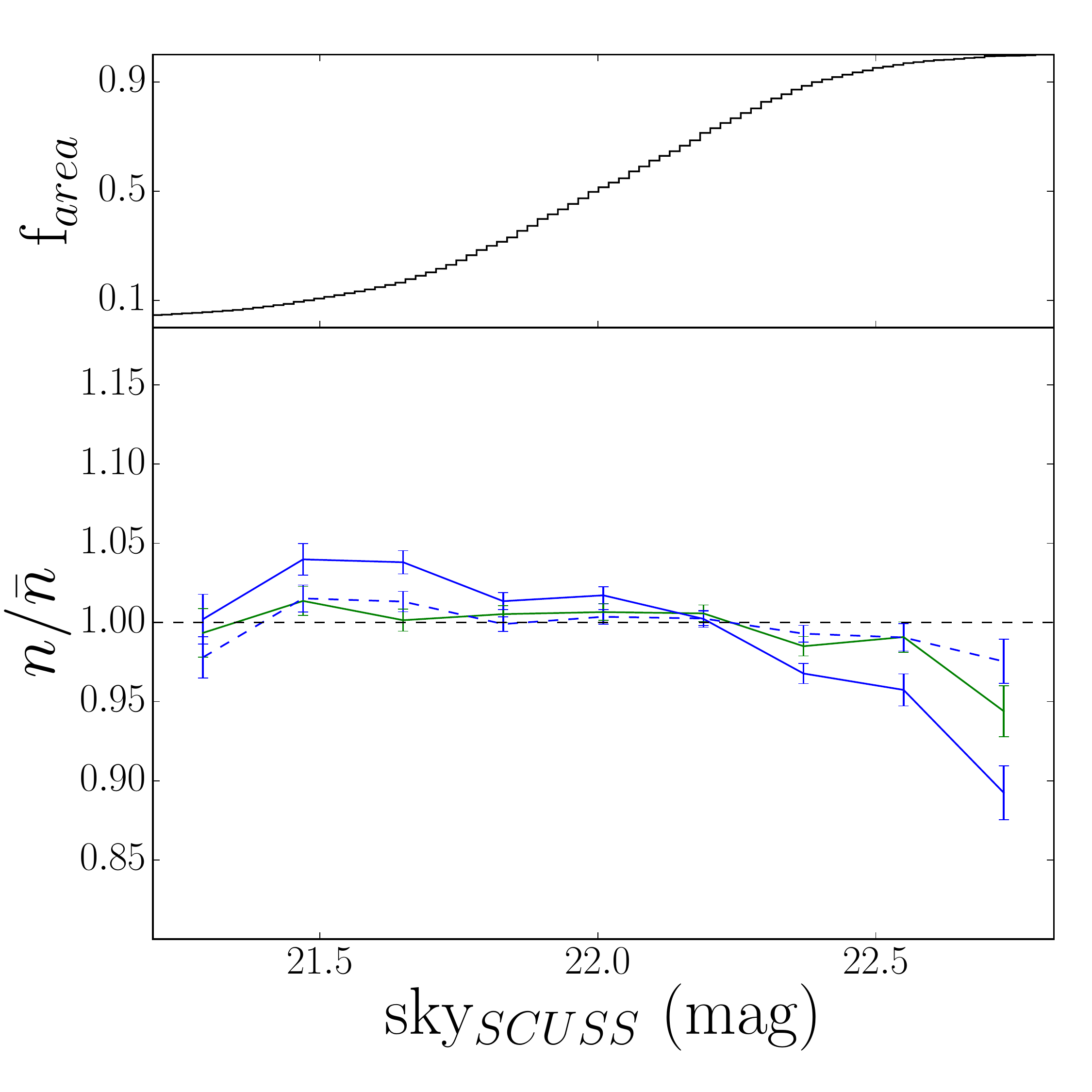,width=5.3cm}

\caption[]{Same as Fig.~\ref{fig:systmaps} but for four different observational parameters: SDSS seeing (seeing\textsubscript{SDSS}), WISE $W1$ median coverage (covmedian), SCUSS seeing (seeing\textsubscript{SCUSS}) and SCUSS sky flux (sky\textsubscript{SCUSS}).}
\label{fig:systmaps2}
\end{center}
\end{figure*}

We estimate the value of the observational parameters in the same pixels used to compute the selection densities (i.e \texttt{NSIDE} = 128). The resulting maps are displayed in Fig.~\ref{fig:systmaps} and~\ref{fig:systmaps2}. Those figures also show the evolution of the normalised number density $(n/\bar{n})$ as a function of the different observational parameters for the two selections. Both the griW and UgrizW normalised number densities exhibit similar non-linear dependencies on stellar density, extinction and SDSS sky flux. To contain the propagation of errors due to those parameters, we discard pixels with $n_{star}>6000$, $A_g > 0.5$, and sky\textsubscript{SDSS}~$>2.4$, removing a total of 12\% of the initial area.

The stellar density $n_{stars}$ has a strong systematic effect on both selections, with the selection densities dropping by 15 to 20\% when the stellar density increases from 1000 to 5000\,\perdegsq. One usually expects the density of the selection to increase with increasing stellar density due to the stellar contamination of the sample (i.e. stars being erroneously selected as potential ELGs).
Actually, as shown by~\citet{Ross11} (in the following ROSS11), the stellar density field also has an opposite effect: due to observational effects (including the seeing and the point spread function of the telescope) each star is observed as a slightly extended object, masking a small fraction of the sky and preventing the selection of galaxies in that region. The negative slope seen in Fig.~\ref{fig:systmaps} indicates that this latter effect is dominant for both selections. We conduct a quantitative study of the area masked by foreground stars in Sec~\ref{sec:Stars}.

The Galactic extinction $A_g$ also has a strong effect on the UgrizW selection density, which decreases by about 20\% as the extinction increases from 0.05 to 0.35 magnitude. The effect of the extinction on the griW selection is weaker: it is contained within 8\% over the same range. This effect is expected as over high extinction regions, the observed flux of our targets is closer to the limiting magnitude of detection, making them harder to select. 

The sky\textsubscript{SDSS} has a relatively weak effect on both selections, with the fluctuations in number densities being contained below 8\% for the griW selection and below 5\% for the UgrizW, between 1.3 and 2.0\,nanomaggies.arcsec$^{-2}$ which corresponds to 80\% of our masked footprint.  

The systematic effects induced by the airmass\textsubscript{SDSS} are comparable to the effects linked to the sky\textsubscript{SDSS} as the two maps are highly correlated. The effect on the griW selection corresponds to a linear increase of about 7\% over the range 1.1 to 1.7; this corresponds to 85\% of our masked footprint. The airmass\textsubscript{SDSS} has essentially no effect on the UgrizW selection density. 

\begin{figure}
\begin{center}
\epsfig{figure=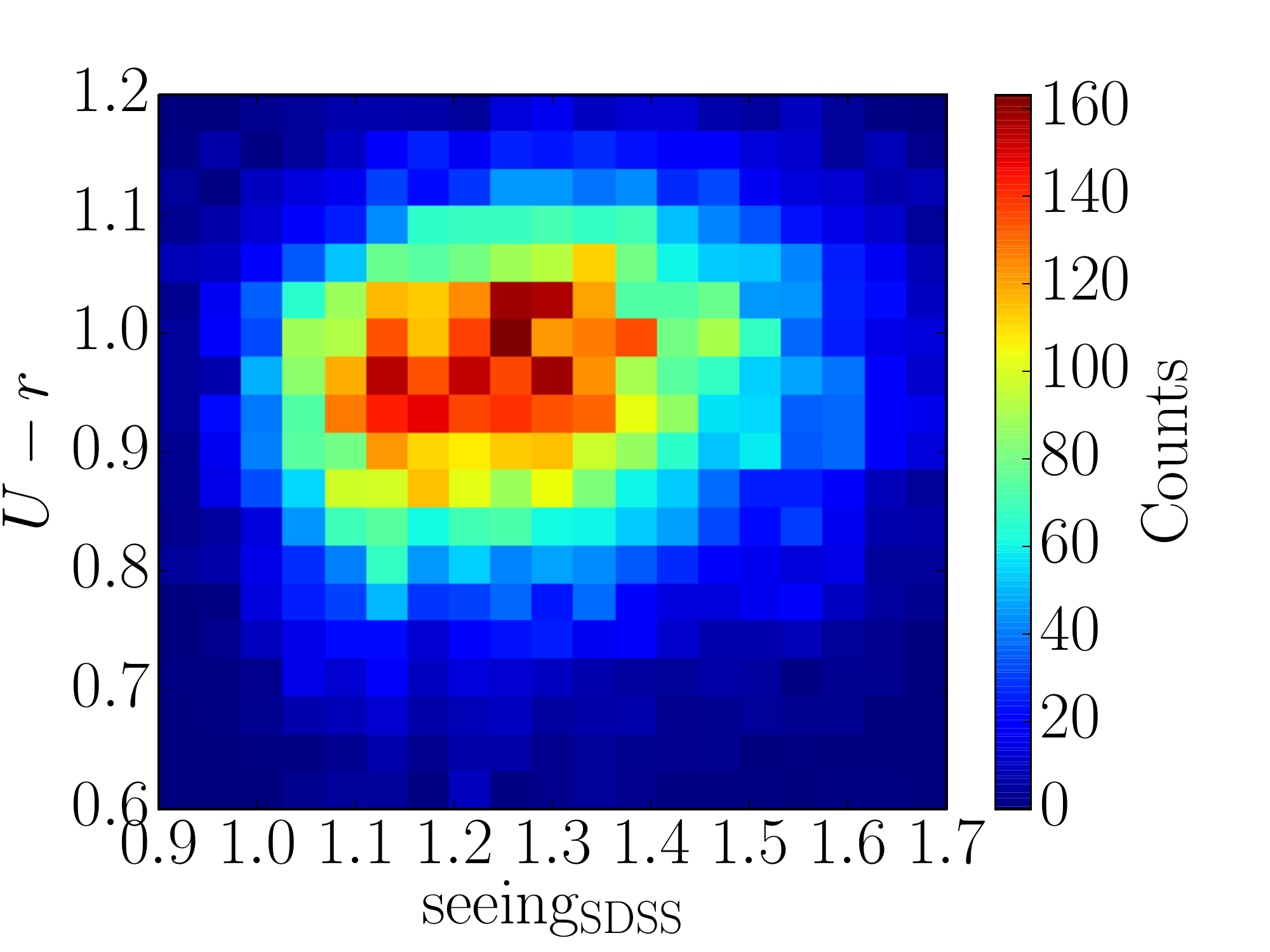,width=7cm}
\epsfig{figure=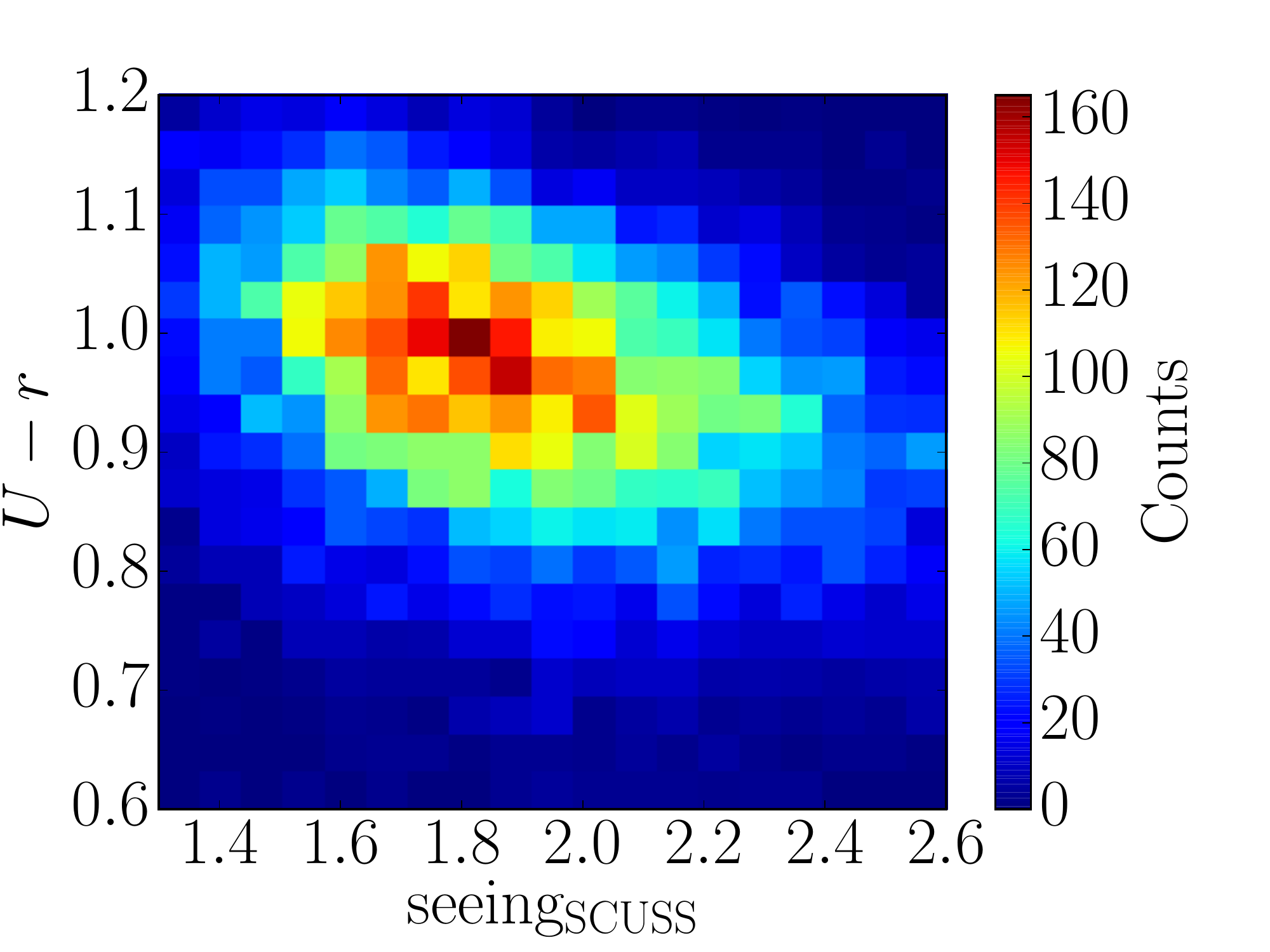,width=7cm}
\epsfig{figure=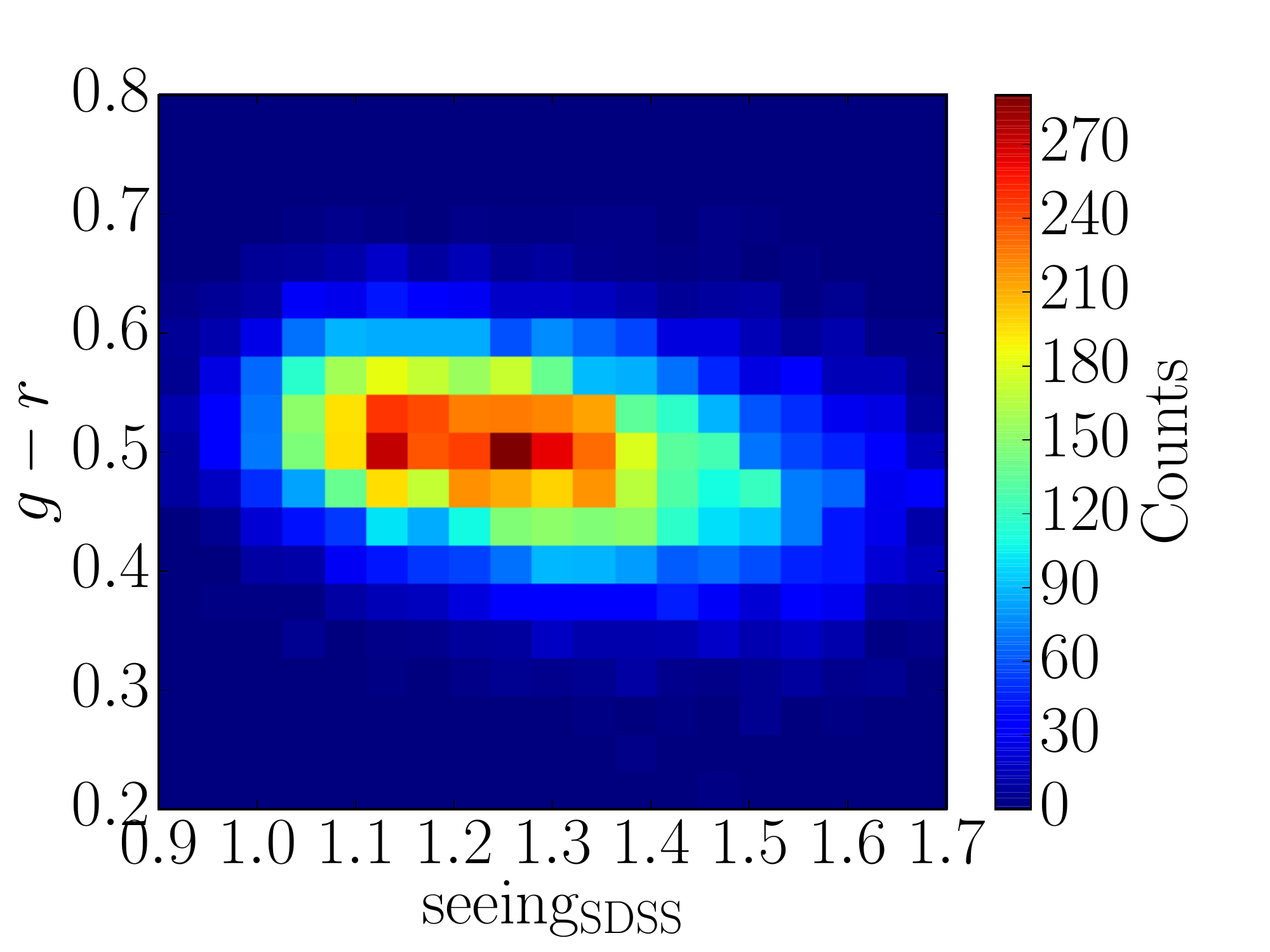,width=7cm}
\caption[]{Two dimensional histograms showing the distribution of the pixels of the UgrizW selection as a function of the mean values of the $U - r$ color and of the seeing\textsubscript{SDSS} (top panel) or seeing\textsubscript{SCUSS} (middle panel) in the pixel. The correlations between the color and the values of the seeings are clearly visible. For comparison we also present the distribution of the pixels of the UgrizW selection as a function of the $g - r$ color and of the seeing\textsubscript{SDSS} (bottom panel) that shows very few correlations.}
\label{fig:seeing}
\end{center}
\end{figure}

The impact of the seeing is quite different on the two selections. The variation of densities of the griW selection as a function of SDSS seeing is contained within 5\% over the full masked footprint. As expected, there is no variation as a function of SCUSS seeing. The UgrizW selection density, however, exhibits a strong correlation with both SDSS and SCUSS seeings. The selection density drops by 20\% over the range 1.05 to 1.6 arcsec, which includes more than 90\% of the survey. The correlation with SCUSS seeing is opposite and even stronger as the density of the selection increases by 30\% over the range 1.2 to 2.8. This tight correlation between the UgrizW selection density and SDSS and SCUSS seeing is due to correlations between those seeing parameters and the color of objects in our catalog. Figure~\ref{fig:seeing} presents the distribution of the pixels in two dimensional histograms with the mean $U-r$ color of the objects in a pixel on the $y$-axis and the seeing on the $x$-axis. The $U-r$ color term has a positive correlation with SDSS seeing, where as it has a negative correlation with SCUSS seeing. This color term enters the Fisher discriminant $X_{FI}$ of Eq.~\ref{eq:Fisher} multiplied by a negative parameter given in Tab.~\ref{tab:Fisher}. A higher value of SDSS seeing implies on average a higher value of $U-r$, thus a lower value of $X_{FI}$: less objects will be selected. Conversely, a higher value of SCUSS seeing produces on average a lower value of $U-r$, thus a higher value of $X_{FI}$: more objects will be selected.

The UgrizW selection has a 15\% variation of density as a function of WISE median coverage over the range 22 to 40 that includes more than 95\% of our masked footprint. The effect of covmedian is smaller on the griW selection density as the effect is contained within 10\% over the full range of values of the observational parameter. 

The UgrizW selection density is slightly anti-correlated with SCUSS sky flux, exhibiting a 10\% decrease over the range 21.2 to 22.5, which covers 90\% of the footprint. 

\subsection{Foreground stars}\label{sec:Stars}

\begin{figure}
\begin{center}
\epsfig{figure=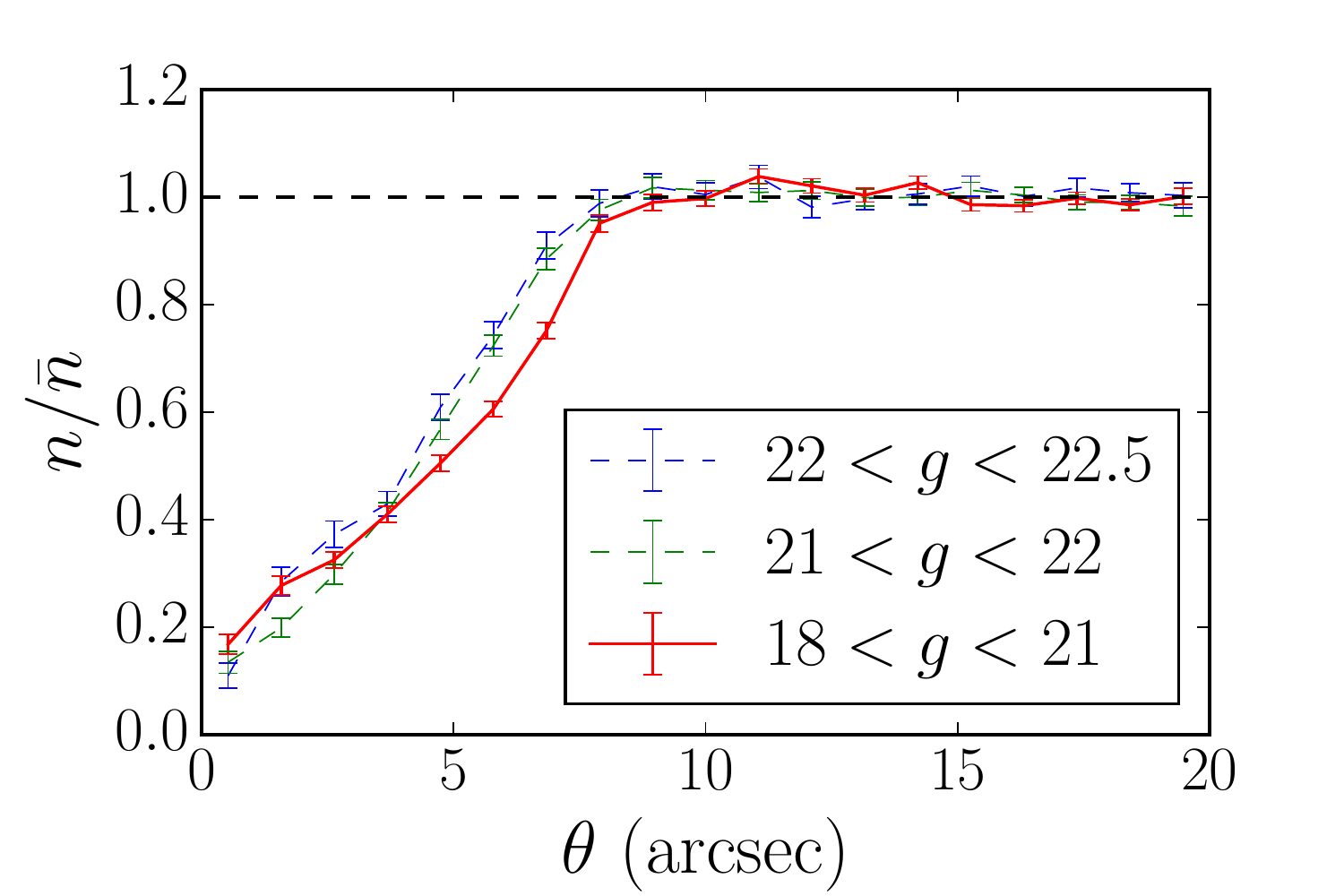,width=9cm}
\caption[]{Number density of the griW selection normalised to the average number density of the selection in annuli of 1 arcsec width centered around stars as a function of the radius of the annulus. Our star sample corresponds to the $18<g<21$ curve. For comparison, we also plot the curves for samples selected in the same manner as our stellar sample but with $21<g<22$ and $22<g<22.5$. For each sample there is a clear decrement of targets for radii below 8 arcsec.}
\label{fig:aroundstars}
\end{center}
\end{figure}

To quantify the area masked by each star in our star sample we estimate the number density of objects in our selections in annuli of different radii and of 1 arcsec width surrounding stars in our masked stellar map. We then obtain the mean number density around stars by averaging over all the stars in our sample. The resulting number density, normalised to the mean number density of our selection as a function of the radius of the annulus, is displayed in Fig.~\ref{fig:aroundstars} for the griW selection. There is a clear deficit of targets surrounding stars at separations below 8 arcsec, while above that radius the number of targets surrounding stars is consistent with the average of the selection. We compute the effective area masked by each star $\mathcal{A}_{masked}$ as 
\begin{equation}
\mathcal{A}_{masked} = \int 2\pi\left(1-\frac{n}{\bar{n}}\right)\theta \mathrm{d}\theta,
\end{equation}
where $\theta$ is the radius of the annulus and $n/\bar{n}$ is taken from Fig.~\ref{fig:aroundstars}. We obtain $\mathcal{A}_{masked} = 78.0$~arcsec$^2$ for the griW selection and $\mathcal{A}_{masked} = 100.1$~arcsec$^2$ for the UgrizW selection.

Given a complete and pure sample of stars that mask a part of our footprint, we could compensate for this effect by correcting the number densities of the selections in each pixel given the stellar density in the pixel.  As we select only bright ($18<g<21$) point sources, we are confident that the purity of our stellar sample is high. However, Fig.~\ref{fig:aroundstars} shows that point sources with $21<g<22$ and $22<g<22.5$ mask a similar area of the sky than stars of our stellar sample. Selecting point sources at magnitude 22 and fainter is not a satisfactory approach to construct a pure stellar sample as the contamination by non-extended galaxies becomes non negligible. Thus, we do not apply the correction and proceed by modelling the dependency on the stellar density of our selections, together with other parameters, as described in the next section. 

\subsection{Modelling the systematic effects}\label{sec:modelsyst}

We follow the procedure of \citet{Prakash16} to model the effect of the observational parameters on the number density of the selections. We assume that the observed number density in pixel $p$ can be expressed as a function of all the parameters by
\begin{equation}\label{eq:syst_model}
n_{p} = \bar{n}\left[1 + \sum_i\sum_ka^k_i\left(\frac{s_p^i - \tilde{s}_i}{\Delta s_i}\right)^k\right]+ \epsilon_p ,
\end{equation}
where $\bar{n}$ corresponds to the average number density, $s_i$ the value of the parameter $i$, $\tilde{s}_i$ the median value of the parameter $i$ over the footprint, $\Delta s_i = \rm{max}(s_i) - \rm{min}(s_i)$ and $\epsilon_p$ is a term accounting for shot noise and cosmic variance. We have assumed that the effect of each parameter $i$ on the observed number density can be modelled as a polynomial function of the value of the parameter with coefficients $a^k_i$, and that the shot noise and cosmic variance are Gaussian. The shot noise assumption is insured by the fact that we use sufficiently wide pixels such that the mean number of entry per pixel is $\approx37$ for the griW selection and $\approx42$ for the UgrizW one.
We estimate the value of all coefficients $a^k_i$ using a multivariate regression. The $a^0_i$ parameters are degenerate and therefore set to 0. For each observational parameter, we limit ourselves to first ($k=1$) or second order ($k\leq2$) polynomial depending on whether the density shows obvious non-linear dependency on the parameter. The orders of the polynomials fitted for the different parameters are listed in Tab.~\ref{tab:syst}.

Knowing the value of the different observational parameters for each pixel, we can compute a predicted density per pixel
\begin{equation}\label{eq:syst_pred}
n_{p}^{pred} = \bar{n}\left[1 + \sum_i\sum_ka^k_i\left(\frac{s_p^i - \tilde{s}_i}{\Delta s_i}\right)^k\right],
\end{equation}
where, compared to Eq.~\ref{eq:syst_model}, we have now removed the shot noise and cosmic variance. The middle panels of Fig.~\ref{fig:predicteddens} shows the predicted density maps for the two selections. As expected, the predicted density maps exhibit some of the characteristics of the dominant maps of the observational parameters. For instance the unmasked high extinction region near RA of -30$\degree$ and Dec of -2$\degree$ creates an over-density of targets in both selections. Also visible on the griW predicted density map is a general trend of having lower densities at lower RA corresponding to an increase in stellar density as we approach the Galactic plane. The large stripes resulting from the SDSS time-delay imaging strategy, visible on the sky\textsubscript{SDSS} map, are also present on the predicted density maps, while the patchy structure of the seeing\textsubscript{SCUSS} map appears on the UgrizW predicted density map.

\begin{figure}
\begin{center}
\epsfig{figure=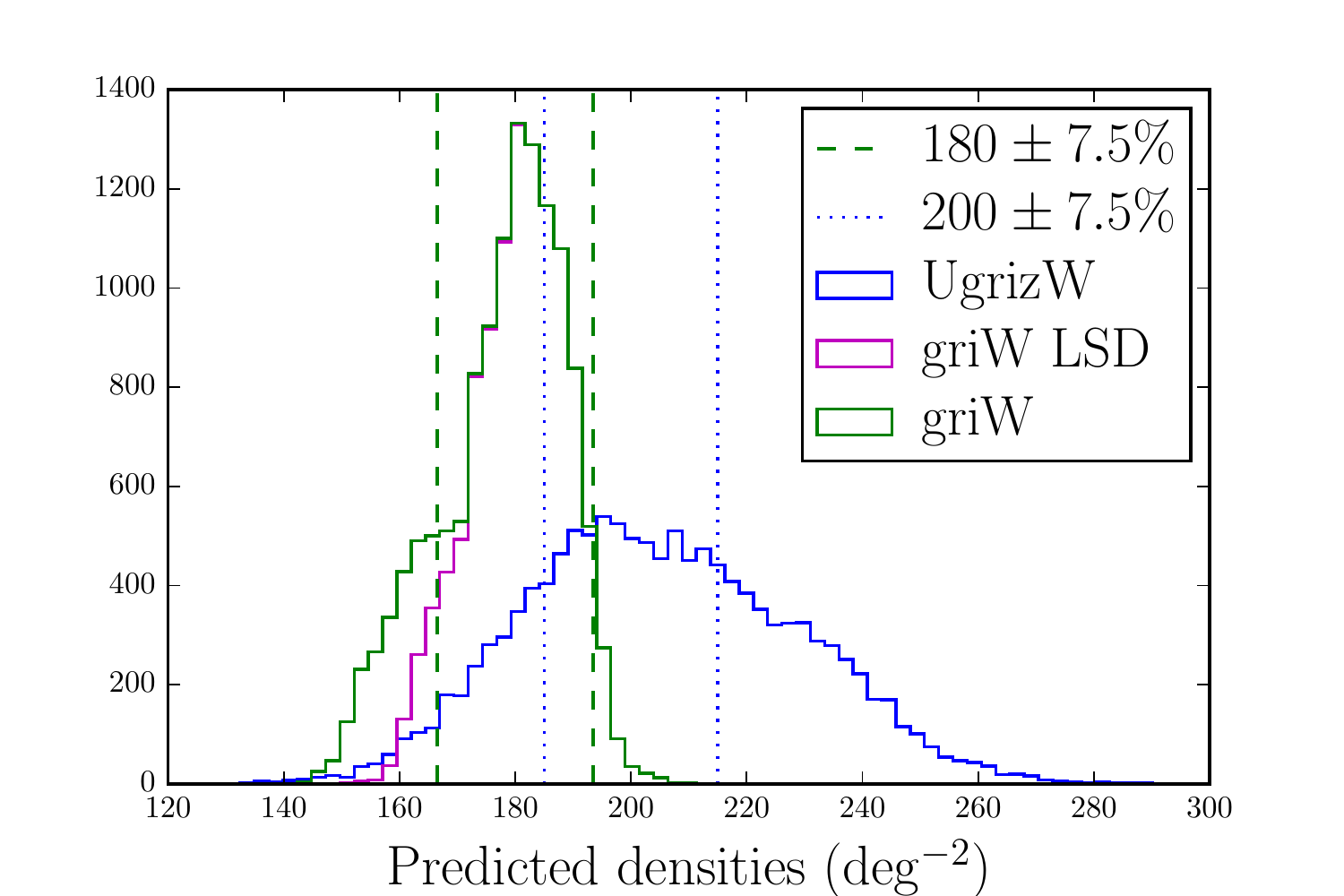,width=9cm}
\caption{Distribution of the pixel densities for both selections. The griW selection is much more homogeneous than the UgrizW one.  For indication, we trace two regions corresponding to fluctuations of $\pm 7.5\%$ around 180\,\perdegsq~and 200\,\perdegsq~which is the total variation allowed by the requirements of Sec.~\ref{sec:requirements}. The mean densities of 180\,\perdegsq~and 200\,\perdegsq~are somewhat arbitrary but roughly correspond to the peaks of the two distributions. For the peculiar case of the griW selection we also plot in dotted line the distribution when only considering Low Stellar Density (LSD) regions (defined as $n_{stars} < 3000$) showing that the entire low density tail is due to high stellar density regions.}
\label{fig:preddenshist}
\end{center}
\end{figure}

The distribution of the pixel densities are visible on Fig.~\ref{fig:preddenshist}. The distribution corresponding to the griW selection is much narrower than that of the UgrizW selection, which shows that this latter selection is more impacted by systematic effects. We emphasise two regions corresponding to a $\pm7.5$\% fluctuation around 180\,\perdegsq~and 200\,\perdegsq~respectively. The choice of the central values is somewhat arbitrary, but they roughly correspond to the maximum of the distribution of the two selections, i.e., they are close to maximising the surface of the footprint passing the homogeneity requirement stated in Sec.~\ref{sec:requirements}. With these central values, 2121\,\degsq~of the griW map pass the homogeneity requirement where as only 1242\,\degsq~of the UgrizW selection do. Thus it is possible to define a 1500\,\degsq~footprint passing the requirements using the griW selection but not the UgrizW one. The tail of the distribution of the griW selection on Fig.~\ref{fig:preddenshist} toward the low densities is only due to the masking effect of stars studied in Sec.~\ref{sec:Stars}; it can simply be removed by applying a tighter cut on the stellar density such as $n_{stars} < 3000$.

\subsection{Correcting for systematic effects}\label{sec:weights}

We have investigated the effects of six observational parameters on the griW selection and eight on the UgrizW selection. We have identified that the major source of potential systematic effects for both selections arise from the stellar density, sky\textsubscript{SDSS}, and the extinction $A_g$, whereas the UgrizW selection is also affected by both SDSS and SCUSS seeings. To limit the influence of those observational parameters we have removed a small portion of the footprint that contains the most extreme values of the parameters. 
 
As shown by ROSS11, we can further reduce the effect of those observational parameters by defining an appropriate weighting procedure. Given Eq.~\ref{eq:syst_model}, an obvious expression for the weights is given by
\begin{equation}\label{eq:weights}
w_{p} = 1+\frac{\bar{n}-n_p^{pred}}{n_p}.
\end{equation}
Applying those weights to the pixel densities reduces the systematic effects, as shown on Fig.~\ref{fig:systmaps} and~\ref{fig:systmaps2}, where the fluctuations of the reduced average number density are now consistent with zero given the uncertainties computed as the root mean square (RMS) in the bin. 

\subsection{Zero-point fluctuations}

\begin{table*}
\begin{center}
\begin{tabular}{
	l
	S[table-format=3.2]
	S[table-format=3.2]
	S[table-format=3.2]
	S[table-format=3.2]
	S[table-format=3.2]
	}
\toprule
photometric & \multicolumn{1}{c}{error} & \multicolumn{2}{c}{\emph{UgrizW}} &  \multicolumn{2}{c}{\emph{griW}} \\

band & \multicolumn{1}{c}{(mmag)}& \multicolumn{1}{c}{$\frac{1}{N}\frac{\Delta N}{\Delta m}$} & \multicolumn{1}{c}{fluctuation (\%)} & \multicolumn{1}{c}{$\frac{1}{N}\frac{\Delta N}{\Delta m}$} & \multicolumn{1}{c}{fluctuation (\%)} \\
\midrule
SCUSS \emph{U} &22&-1.1&9.4& \multicolumn{1}{c}{-} & \multicolumn{1}{c}{-} \\
SDSS \emph{g} &9&-2.4& 8.6&-3.6&13\\
SDSS \emph{r} &7&3.5& 9.8&4.3&12\\
SDSS \emph{i} &7&-0.7& 1.9&-1.7&4.8\\
SDSS \emph{z} &8&-0.3& 1.0& \multicolumn{1}{c}{-} & \multicolumn{1}{c}{-} \\
WISE \emph{W1} &16&-1.1&7.3&-1.2&7.8\\
\bottomrule
\end{tabular}
\caption{The impact of fluctuations in imaging zero-points on the number densities of the selections. The \emph{error} column lists the estimated $1\sigma$ error on the zero-point for the different photometric bands in mmag. For each selection, the first column corresponds to the normalised variation in number density due to a variation of the zero-point $\Delta m$, while the second column gives the resulting variation of number density in percent over 95\% of the footprint.}
\label{tab:zeropoints}
\end{center}
\end{table*}

The requirements of Sec.~\ref{sec:requirements} also state that the number density of the selection should not vary by more than 15\% over the footprint as a function of imaging zero-points. As in~\citet{Myers15} and~\citet{Prakash16}, we test this requirement by adding $\pm0.01$ mag to each photometric band used and then rerunning the target selection algorithm to estimate the change in target density. We test each photometric band individually. The normalised change in target density due to the shift of a given band ($\frac{1}{N}\frac{\Delta N}{\Delta m}$) is then multiplied by the expected RMS error in the photometric calibration of that band to obtain the expected RMS variation in target density due to shifts of the imaging zero-point. We use the $1\sigma$ error estimates of~\citet{Finkbeiner16} for the SDSS bands (summarised in Tab.~\ref{tab:zeropoints}) and estimate a $1\sigma$ error of 0.016 for $W1$ from~\citet{Jarrett11}. We adopt a SCUSS zero-point $1\sigma$ error of the order of 22\,mmag from private communication with the SCUSS collaboration. 

Assuming Gaussian errors, 95\% of our footprint lies within a $\pm2\sigma$ variation from the expected zero-point of any given photometric band, meaning that 95\% of our footprint has a variation in target number density lower than $4\times \sigma_{zp} \times\frac{1}{N}\frac{\Delta N}{\Delta m}$, where $\sigma_{zp}$ is the RMS error on the zero-point in the relevant photometric band. The resulting fluctuations for each photometric band are given in Tab.~\ref{tab:zeropoints}. The strongest fluctuations are obtained with SCUSS $U$-band and  SDSS $r$-band on the UgrizW selection. However, those fluctuations are below 10\% over 95\% of the footprint, below the 15\% requirements. For the griW selection, the dominant fluctuations are due to $g$-band and $r$-band with 13\% and 12\% respectively. Thus both selections are robust against variation of the imaging zero-points.

\section{Large scale Angular Clustering of pixels}\label{sec:clustering}

\subsection{Method and model}

Following~\citet{Scranton02} and ROSS11, we start by computing the angular clustering directly using the pixels defined in the previous section. For each pixel $p$ we define the galaxy over-density as 
\begin{equation}
\delta^g_p = \frac{n_p}{\bar{n}} - 1,
\end{equation}
where $n_p$ is the number density of galaxies in the pixel and $\bar{n}$ the mean number density over all the pixels. We also compute the fluctuations with respect to the mean for each observational parameter $s_i$ as
\begin{equation}
\delta^{s_i}_p = \frac{s_p^i}{\bar{s}_i} - 1,
\end{equation}
where $s_p^i$ is the value of the observational parameter $i$ in the pixel $p$ and $\bar{s}_i$ is the mean value of the parameter over all the pixels. We can therefore compute the correlation function as 
\begin{equation}
w(\theta) = \frac{\sum_{p,q}\delta^{\alpha}_p\delta^{\beta}_q\Theta_{p,q}}{\sum_{p,q}\Theta_{p,q}},
\end{equation}
where $\Theta_{p,q}$ equals 1 if the angular separation between pixels $p$ and $q$ falls into the angular bin $\theta$ and zero otherwise, and $\alpha$ and $\beta$ denote either the galaxy over-density $g$ or an observational parameter $s_i$. Thus the previous equation corresponds to an auto-correlation in case $\alpha=\beta$, and a cross-correlation in case $\alpha\neq\beta$. Computing the angular clustering directly with the pixels has the advantages of being much faster than doing the full computation with the individual galaxies and does not require the use of randoms as we are dealing with continuous fields. We are limited by the resolution of the pixels $\theta_{\rm{pix}}\sim 0.46$\,deg. 

We compute the errors on the angular clustering using a Jackknife estimator~(see e.g. ROSS11)
\begin{equation}
\sigma^2_{Jack}(\theta) = \frac{N_s - 1}{N_s}\sum_{i=1}^{N_s}\left[w(\theta) - w_i(\theta)\right]^2,
\end{equation}
where $N_s$ is the number of subsamples, $w$ is the angular correlation over the full sample, and $w_i$ is the angular correlation over the subsample $i$. We divide our masked footprint into 24 equal area subsamples by searching for continuous regions of neighbouring pixels with surface of $\mathcal{A}_{tot}/N_s$, where $\mathcal{A}_{tot}$ is the area of the total footprint. 

\subsection{Measurements}

\begin{figure}
\begin{center}
\epsfig{figure=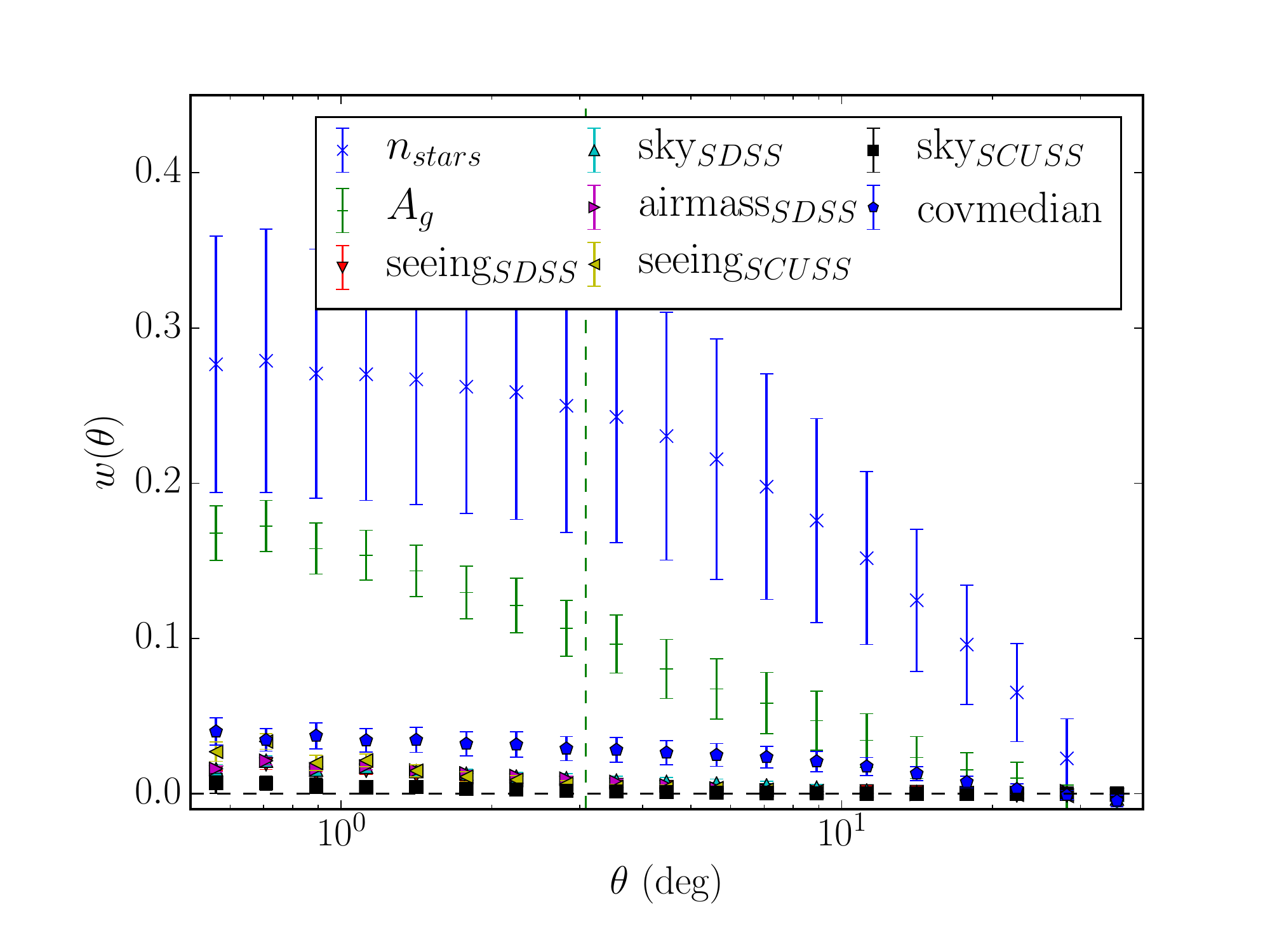,width=8cm}
\caption{Auto-correlation of the eight observational parameters considered. The parameters with the strongest signals are the stellar density and the extinction $A_g$, thus they are the most likely to induce errors on the clustering measurement. The signals in airmass, sky\textsubscript{SDSS}, seeing\textsubscript{SDSS}, and seeing\textsubscript{SCUSS} are comparable. The error bars are computed using a Jackknife estimator. The vertical green dashed line corresponds to the expected apparent BAO scale at $z=0.76$.}
\label{fig:autocorr}
\end{center}
\end{figure}

Figure~\ref{fig:autocorr} shows the auto-correlations of the different observational parameters we consider. Consistently with ROSS11, the strongest signal comes from the stellar density auto-correlation with a value of $0.23\pm0.08$ at an angular separation $\theta_{\rm{BAO}}=3.0$\,deg which corresponds to the expected BAO scale at redshift 0.8. The quoted error is the $1\sigma$ diagonal error on the correlation measurement. The second dominant signal is the extinction $A_g$ that is about two times lower than the stellar density auto-correlation signal, with a value of $0.10\pm0.02$ at $\theta_{\rm{BAO}}$. All of the remaining observational parameters exhibit much lower signal; airmass, sky\textsubscript{SDSS}, seeing\textsubscript{SDSS}, and seeing\textsubscript{SCUSS} have similar signals below 0.02 down to the pixel scale $\theta_{\rm{pix}}$, where as WISE covmedian has a slightly stronger signal of $0.03\pm0.008$ at $\theta_{\rm{BAO}}$. The auto-correlation of sky\textsubscript{SCUSS} has the smallest signal with an autocorrelation value of $0.002\pm0.002$ at the BAO scale.

\begin{figure}
\begin{center}
\epsfig{figure=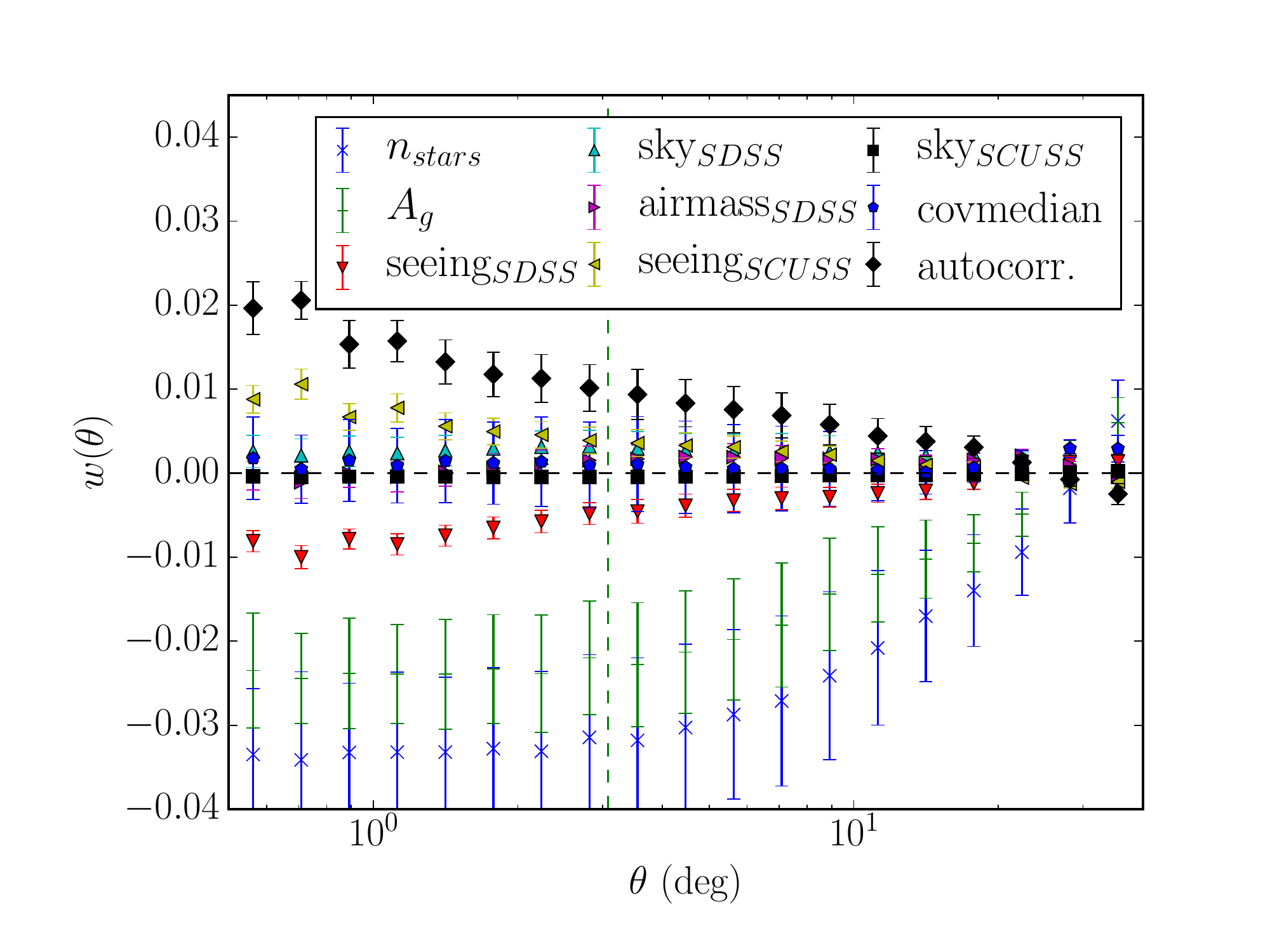,width=8cm}
\epsfig{figure=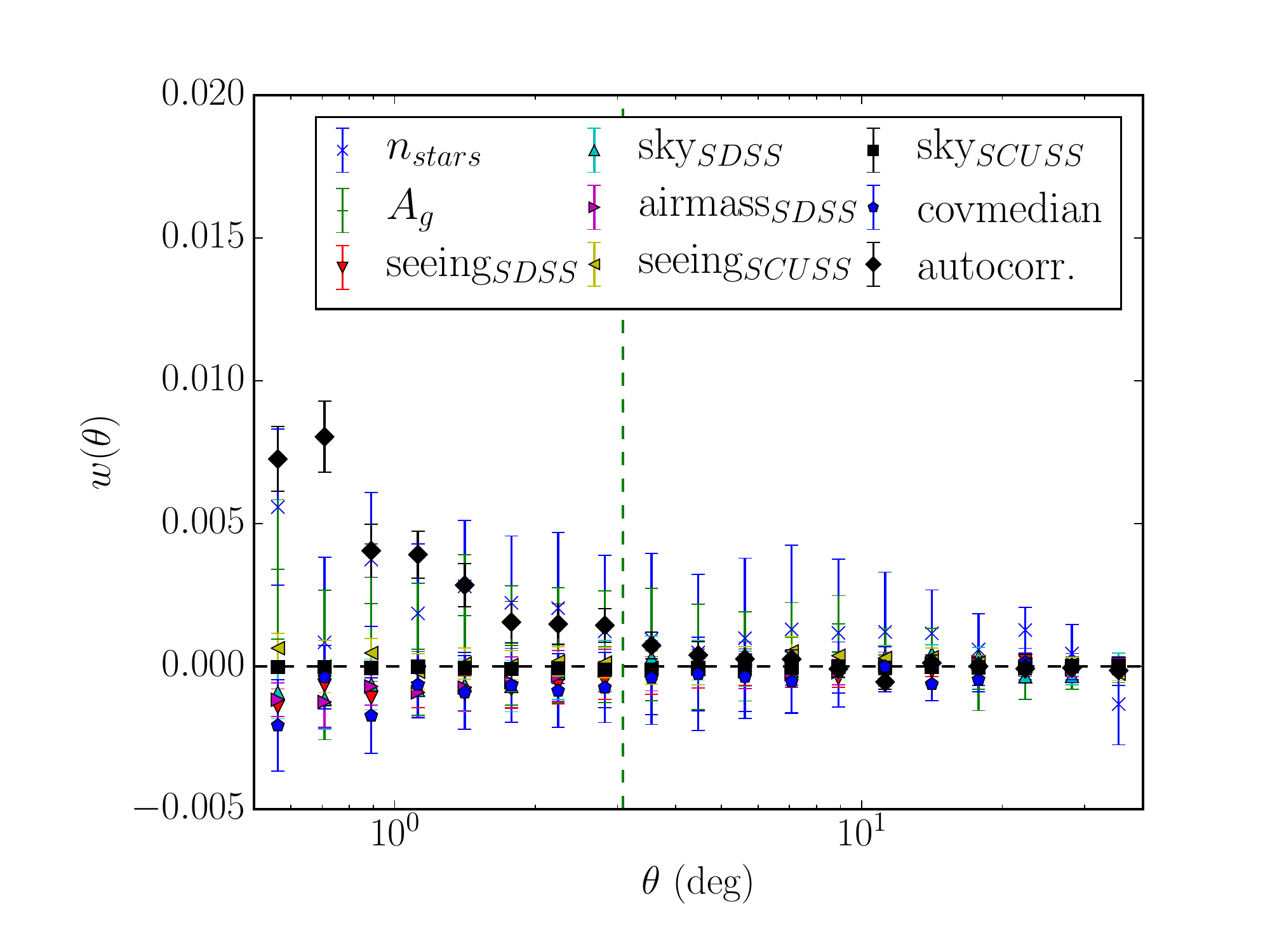,width=8cm}
\caption{Top: Auto-correlation of the galaxy density and cross-correlations between UgrizW selection galaxy density and the different observational parameters. Bottom: Same format after applying the weighting technique. The vertical green dashed line corresponds to the expected apparent BAO scale at $z=0.76$.}
\label{fig:corrUgrizW}
\end{center}
\end{figure}

The angular autocorrelation signal of the UgrizW selection is displayed on the top panel of Fig.~\ref{fig:corrUgrizW}. 

The value of the autocorrelation is $0.015\pm0.003$ at $2\theta_{\rm{pix}}$ and $0.01\pm0.003$ at $\theta_{\rm{BAO}}$. The high amplitude at the BAO scale and above is an indication of the presence of systematic effects. Also visible on this panel is the cross-correlation signal between the selection density and the value of the observational parameters. The cross-correlations with the stellar density and the extinction are larger with respective signals of $-0.031\pm0.01$ and $-0.021\pm0.007$ at the BAO scale. The negative signs are consistent with the negative slopes of the curves showing the evolution of the normalized number densities of the selection as a function of the parameters shown on Fig.~\ref{fig:systmaps}. The next two cross-correlations with the most signal are those involving SCUSS and SDSS seeing, with respective signals of $0.01\pm0.002$ and $-0.01\pm0.001$ on the pixel scale, and smaller signals of $0.004\pm0.002$ and $-0.005\pm0.001$ at the BAO scale. Again the signs of the cross-correlations (i.e., positive with SCUSS seeing and negative with SDSS seeing) are consistent with the trends seen on Fig.~\ref{fig:systmaps2}. All other observational parameters have a cross-correlation with the selection density below $0.003$ in absolute value over the full range of scales. 

The bottom panel of Fig.~\ref{fig:corrUgrizW} presents the angular autocorrelation function of the UgrizW selection density as well as the angular cross-correlations with the observational parameters when applying the correction weights defined in Eq.\ref{eq:weights}. The weighted autocorrelation has a signal of $0.004\pm0.001$ at twice the pixel scale, a third of the value of the unweighted autocorrelation. The value at the BAO scale is $(1.5\pm6)\times10^{-4}$, and the autocorrelation is compatible with zero above the BAO scale, indicating that systematic errors have been removed. This result is confirmed by the cross-correlation measurements, which are now all within $1\sigma$ of zero for separations greater than two pixels. This is a reduction of a factor of 30 for the signal of the cross-correlation with the stellar density at the BAO scale, which is now $0.001\pm0.003$. 

\begin{figure}
\begin{center}
\epsfig{figure=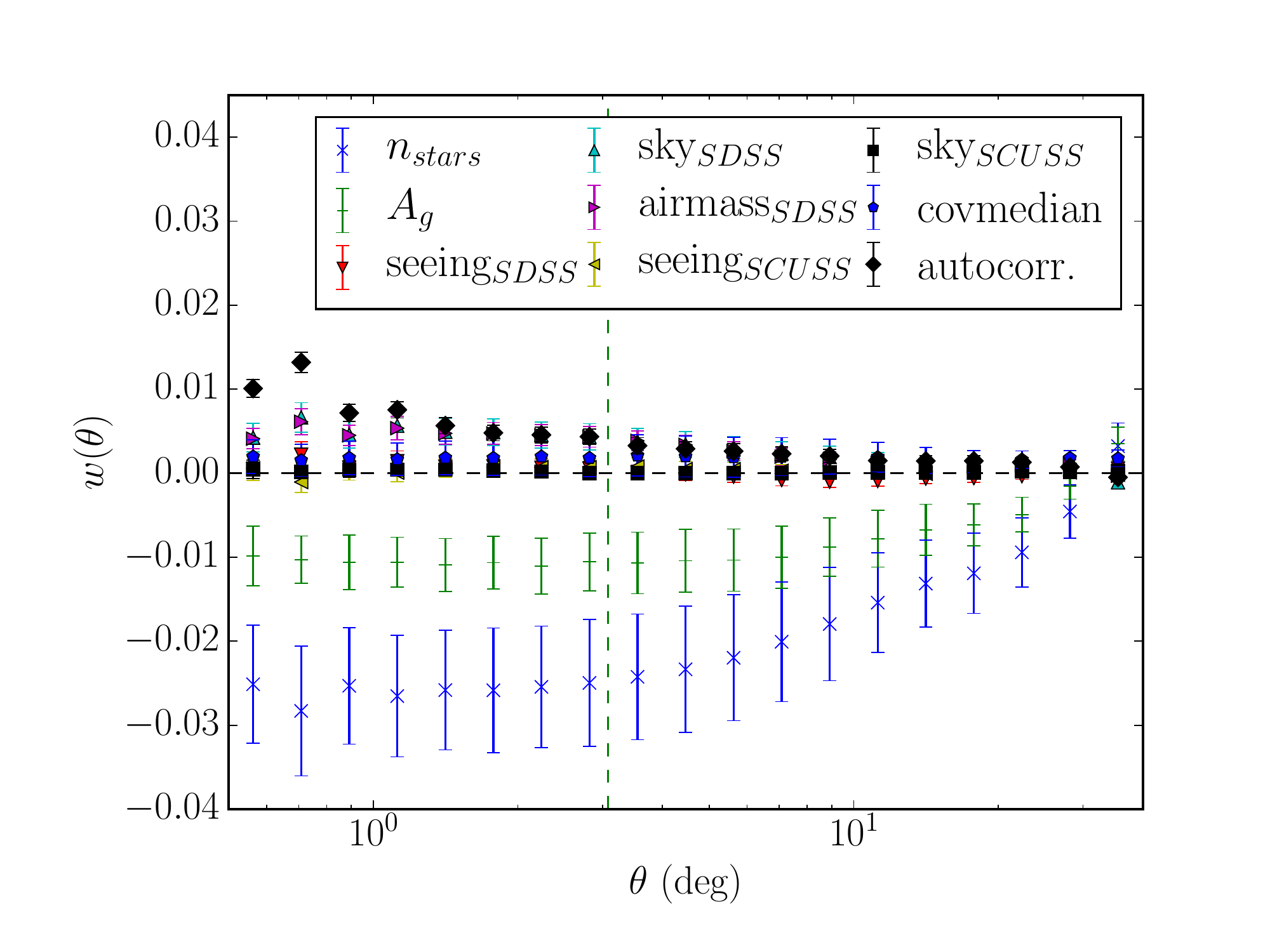,width=8cm}
\epsfig{figure=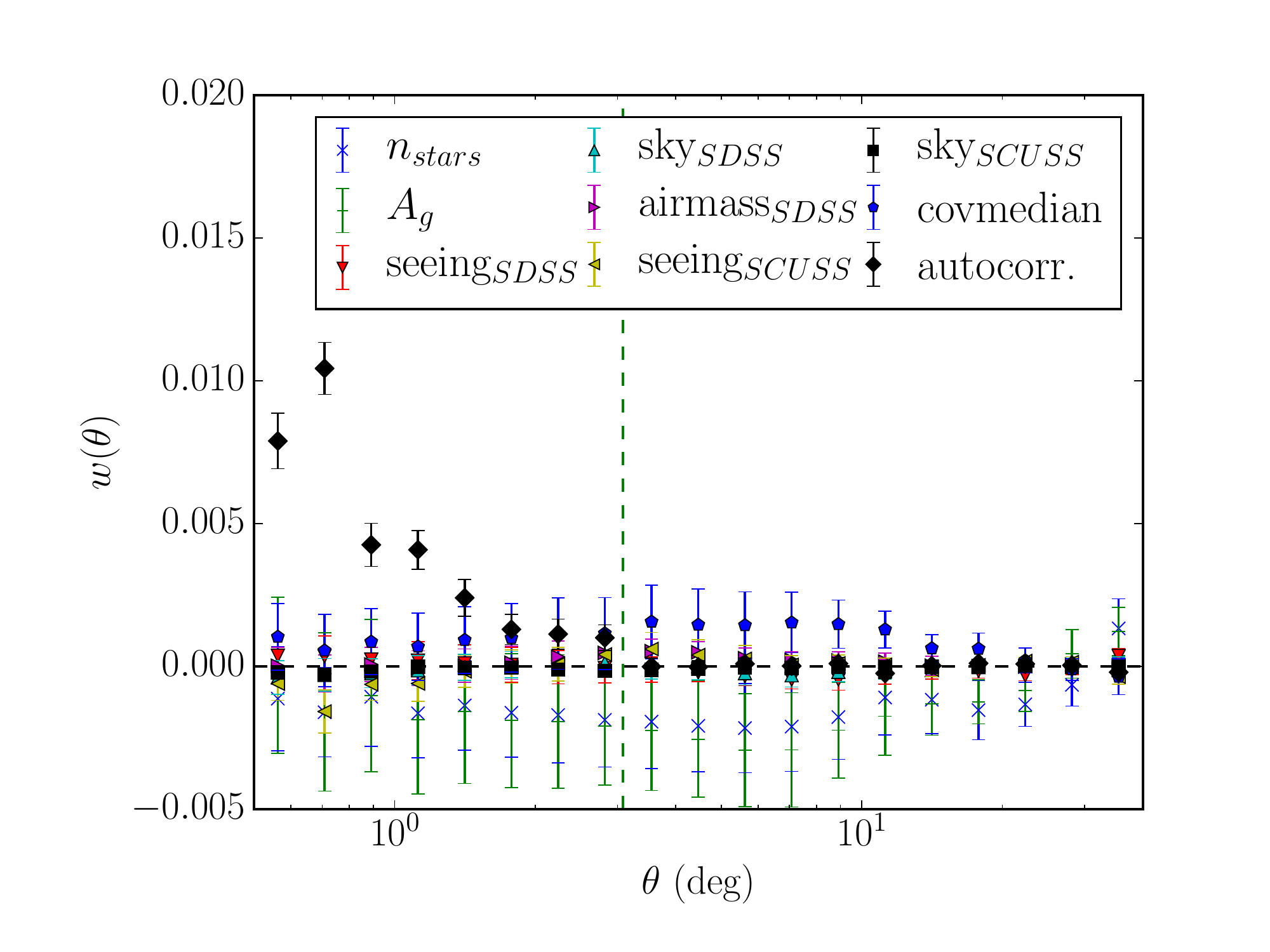,width=8cm}
\caption{Top: Auto-correlation of the galaxy density and cross-correlations between griW selection galaxy density and the different observational parameters. Bottom: Same format after applying the weighting technique. The vertical green dashed line corresponds to the expected apparent BAO scale at $z=0.76$.}
\label{fig:corrgriW}
\end{center}
\end{figure}

The angular autocorrelation signal of the griW selection is presented on the top panel of Fig.~\ref{fig:corrgriW}, together with the cross-correlations between the selection density and the observational parameters. As for the UgrizW selection, the angular correlation function has a high amplitude up to a separation of 20$\degree$, again indicating the presence of systematic errors. The value of the autocorrelation is $0.007\pm0.001$ at $2\theta_{\rm{pix}}$ and $0.004\pm0.001$ at $\theta_{\rm{BAO}}$. The cross-correlation with the stellar density is again the one having the largest signal with a value of $-0.25\pm0.008$ at the BAO scale, slightly smaller than the cross-correlation between the stellar density and the UgrizW selection density. The cross-correlation with the extinction has a value of $-0.01\pm0.003$ at $\theta_{\rm{BAO}}$, roughly half of the cross-correlation between the extinction and the UgrizW selection density. The two cross-correlations with the next higher signals are the ones with the airmass and sky\textsubscript{SDSS}, which both have a signal of $0.004\pm0.001$ at the BAO scale. The cross-correlations with the remaining observational parameters are all below $0.002$ in absolute value at all separations. 

The bottom panel of Fig.~\ref{fig:corrgriW} shows the angular auto and cross-correlation functions for the griW selection when applying the correction weights. As for the UgrizW, there is a drastic  improvement of the angular autocorrelation of the selection density, which now is $(4\pm0.7)\times10^{-3}$ at twice the pixel scale. The value at the BAO scale is $0.001\pm4\times10^{-4}$, and the autocorrelation function is consistent with zero above that scale. As for the UgrizW selection, the cross-correlation signals have been drastically suppressed though the stellar density; the extinction and covmedian stand roughly $1\sigma$ away from zero.  

\begin{figure}
\begin{center}
\epsfig{figure=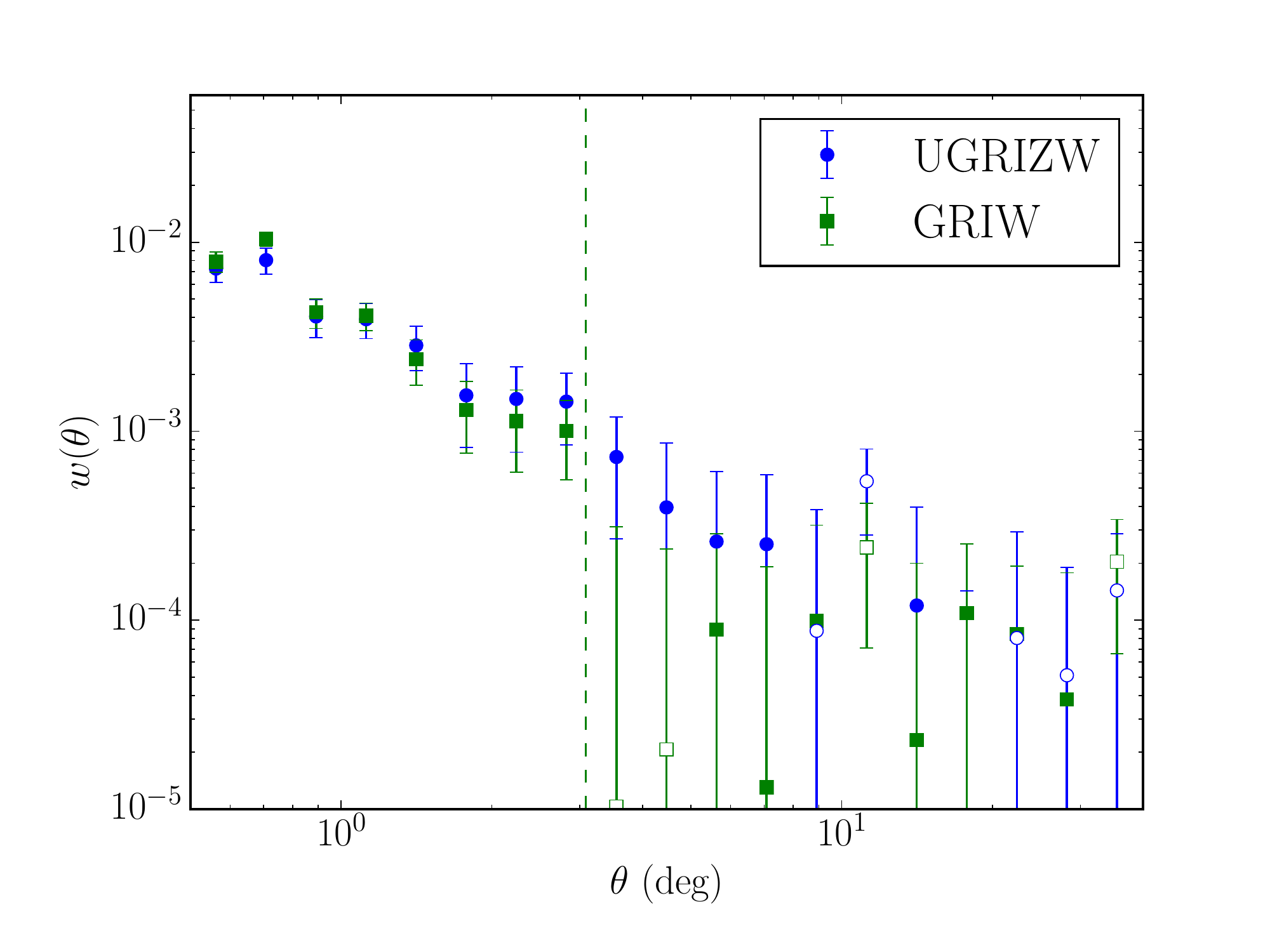,width=8cm}
\caption{Comparison between the UgrizW selection galaxy density auto-correlation and the griW selection one. The vertical green dashed line corresponds to the expected apparent BAO scale at $z=0.76$.}
\label{fig:corr}
\end{center}
\end{figure}

Figure~\ref{fig:corr} compares the two weighted angular autocorrelation functions of the UgrizW and griW selections. Although the two selections have different dependencies on the observational parameters and different unweighted angular autocorrelation functions, the weighted autocorrelations are in good agreement up to the BAO scale. This results indicates that the two selections select similar populations of ELGs.  

\section{Small scale Angular Clustering of galaxies}\label{sec:clusteringgal}

\subsection{Model}\label{sec:wthetamodel}

To model the small scale angular clustering of our catalogs we start with the matter power spectrum $P$ obtained using CAMB\footnote{http://camb.info/}~\citep{Lewis00} and the cosmological parameters defined in Sec.~\ref{sec:intro}. When specified, we include non-linearities using the HALOFIT model~\citep{Takahashi12}. We compute the galaxy distribution power spectrum as
\begin{equation}\label{eq:pk}
P_g(k,z) = b_g(z)^2D_a(z)^2P(k),
\end{equation}
where $D_a$ is the linear growth rate and $b_g$ the bias of our galaxy sample. 
We compute the angular correlation function $w(\theta)$ as an integral of the power spectrum~\citep[e.g.][]{Dodelson02}
\begin{equation}
w(\theta) = \int \mathrm{d}k k^2P(k)W_{\theta}(k),
\end{equation}
where $W_{\theta}$ is the window function defined as 
\begin{equation}
W_{\theta}(k) = \frac{1}{2\pi}\int \mathrm{d}r J_0(kr\theta)n^2(r),
\end{equation}
$J_0$ is the Bessel function, $r$ is the comoving distance and $n(r)$ is the normalised redshift distribution as a function of $r$.

In practice, we compute $P_g$ at the median redshifts of the catalogs given in R16, i.e. $z_{\mathrm{med}}=0.76$ for the griW selection and $z_{\mathrm{med}}=0.78$ for the UgrizW one. When computing $P_g$ including the HALOFIT modelling of non-linearities (Eq.~\ref{eq:pk}), we multiply the matter power spectrum by the cutoff function $\mathrm{exp}(-(0.7k)^2)$ to ensure numerical convergence of the integral. This cutoff does not impact scales greater than $\theta=0.1$\,deg. We use the spectroscopic redshift distributions from R16 to compute $W_{\theta}$.

\subsection{Measurements and bias estimates}

\begin{figure*}
\begin{center}
\epsfig{figure=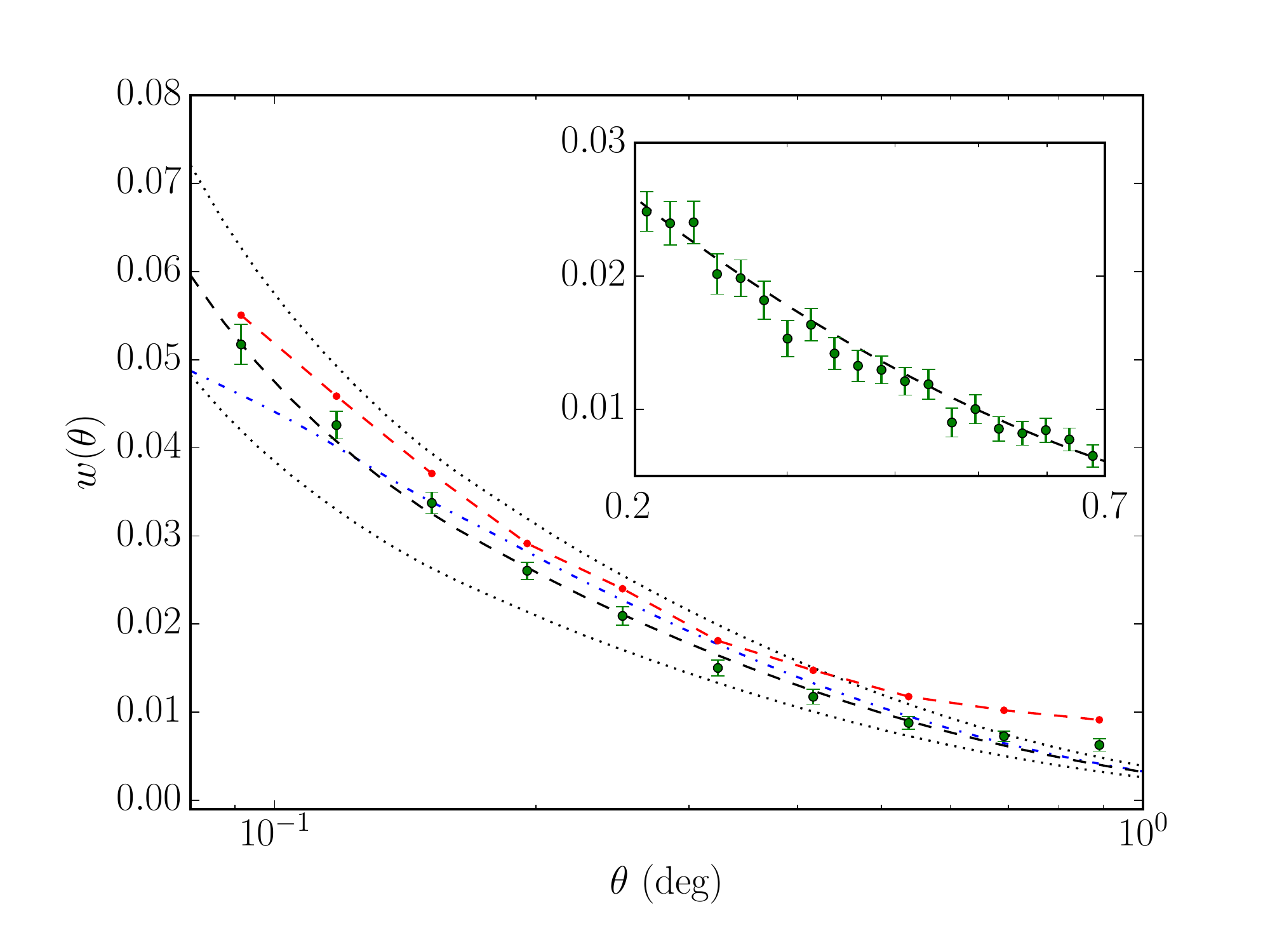,width=8.8cm}
\epsfig{figure=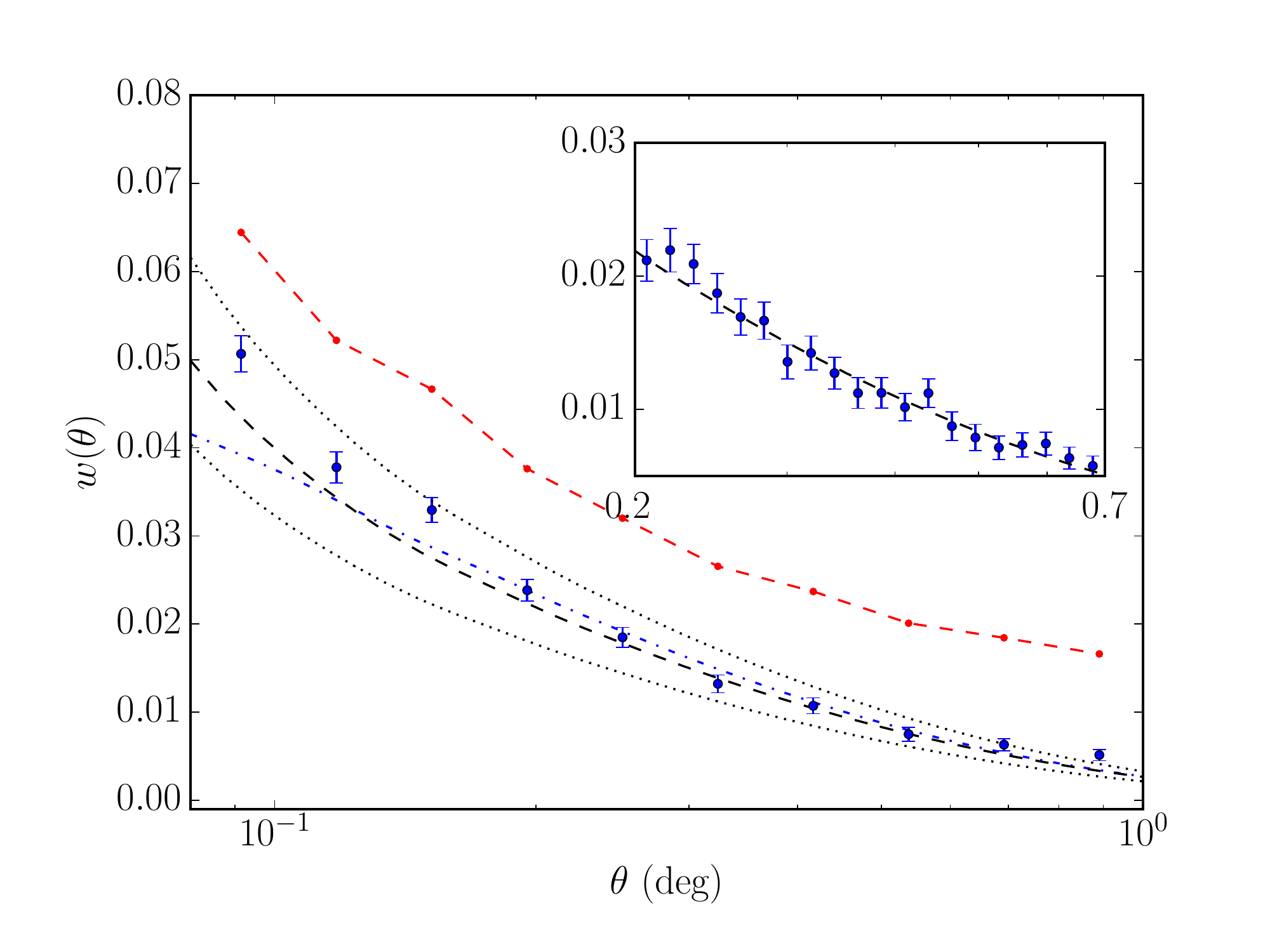,width=8.8cm}
\caption{The angular correlation function of the griW (left) and UgrizW (right) catalogs computed on the discrete objects with the Landy-Szalay estimator. On each plot we show the un-corrected correlation (red dots and dashed line), the corrected correlation (filled green/blue circles), the best fit model as computed in Sec.~\ref{sec:wthetamodel} with HALOFIT correction for the non-linearities (black dashed line) and without it (blue dash-dotted line) as well as two models corresponding to a galaxy bias $b_g=1.8$ and $b_g=2.2$ for the griW plot and $b_g=1.7$ and $b_g=2.1$ for the UgrizW plot respectively (dotted black lines). The inset panels show a blowup of the correlations over the fitting range $0.2<\theta<0.7$ showing the excellent agreement between the measurements and the models over that range.}
\label{fig:wthetagal}
\end{center}
\end{figure*}

We compute the small scale angular clustering of the galaxies using the Landy-Szalay estimator~\citep{Landy93} 
\begin{equation}
w = \frac{DD + RR - 2DR}{RR},
\end{equation}
with and without correcting for the systematic weights. We obtain the covariance matrix using a Jackknife over 128 subsamples of equal number of pairs. Thus the covariance between $w(\theta_i)$ and $w(\theta_j)$ is given by
\begin{equation}
\sigma^2_{Jack}(\theta_i,\theta_j) = \frac{N_s - 1}{N_s}\sum_{p=1}^{N_s}\left[w(\theta_i) - w_p(\theta_i)\right]\left[w(\theta_j) - w_p(\theta_j)\right],
\end{equation} 
where $N_s$ is the number of subsamples, $w(\theta_i)$ is the angular clustering measured over the full catalog at separation $\theta_i$ and  $w_p(\theta_j)$ is the angular clustering measured for the Jackknife subsample $p$ at separation $\theta_j$. The resulting angular correlation functions for both the griW and the UgrizW selections are visible on Fig.~\ref{fig:wthetagal}. We see that the systematic weights remove some clustering at all scales, even lower than the pixel size.

To fit the data we use the theoretical model of Sec.~\ref{sec:wthetamodel} including the non-linearities from HALOFIT, with $b_g$ as the only free parameter. Although the galaxy bias depends on the redshift, we do not fit a parametric function for $b_g$ but rather a single average value for each selection. We fit the measurements over the range $0.2<\theta<0.7$\,deg. The lower bound ensures good modelling of the correlation function while the upper bound removes regions possibly contaminated by remaining systematics. The resulting best fit models are displayed on Fig.~\ref{fig:wthetagal} whereas the best fitted value of $b_g$ together with the $\chi^2$ of the fits are reported in Tab.~\ref{tab:fitbias}. For both the griW and UgrizW catalogs, the resulting best fit models including HALOFIT non-linearities are in very good agreement with the measurements. We obtain a bias of $b_g=2.00\pm0.04$ or $b_g(z) = 1.35/D_a(z)$ for the griW selection and $b_g=1.89\pm0.05$ or $b_g(z)=1.28/D_a(z)$ for the UgrizW one. These results are in agreement with the measurements of~\citet{Comparat13} who found a bias of 1.9 for a similar selection. The slightly higher bias of the griW selection seems to confirm the results of R16 who found that the griW cuts select slightly redder galaxies than the UgrizW ones, thus likely to occupy more massive halos. However we emphasise that the quoted uncertainties are purely statistical and do not include systematics errors which could mitigate the difference. As previously reported (see e.g.~\citet{Crocce16}), the linear theory shows strong discrepancy with the measurements for $\theta<0.1$\,deg.

\begin{table}
\begin{center}
\begin{tabular}{
	c
	c
	c
	}
\toprule
Selection & $b_g$ & $\chi^2/N_{\mathrm{dof}}$ \\
\midrule
griW & $2.00 \pm 0.04$ & 1.28  \\
UgrizW & $1.89\pm0.05$& 1.07\\
 \bottomrule
\end{tabular}
\caption{Best fit value of the galaxy bias $b_g$ and corresponding reduced $\chi^2$ for the two catalogs and for the model of Sec.~\ref{sec:wthetamodel} including HALOFIT non-linearities. The reduced $\chi^2$ are for 19 degrees of freedom.}
\label{tab:fitbias}
\end{center}
\end{table}


\section{Conclusions}\label{sec:conclusions}

eBOSS will provide the first high-precision measurement of the BAO scale using ELGs as a tracer of the matter density field. To reach this goal, eBOSS must obtain spectra of 190,000 confirmed ELGs contained within a footprint of 1500\,\degsq. We have defined two large scale catalogs of ELGs selecting by a Fisher discriminant technique and covering more than 2800\,\degsq~over the SGC. The first catalog, designated UgrizW, uses the $g$, $r$, $i$ and $z$ photometric bands from the SDSS, the $U$ band from SCUSS and the $W1$ band from WISE. The second catalog, designated griW, only uses information from SDSS $g$, $r$ and $i$ photometric bands and WISE $W1$ band. R16 have verified that the selections meet the density, redshift distribution and purity requirements. In this paper we focus on testing the homogeneity requirements. 

We study the dependency of the angular number density of targets in the two selections as a function of observational parameters, including the stellar density, the Galactic extinction, SDSS sky flux, SDSS airmass, SDSS seeing, WISE median coverage, SCUSS seeing and SCUSS sky flux. The angular number densities of both selections depend substantially on the stellar density and the extinction. 
The correlation between the density of both selections and the stellar density is found to be negative, meaning that the larger the density of stars in the given field, the fewer objects are selected. We show that this effect can be understood as resulting from the low stellar contamination of our selections plus the fact that each star masks a small area $\mathcal{A}_{\rm{masked}}$ of the sky, preventing the selection of targets in that area. We estimate $\mathcal{A}_{\rm{masked}}$ to be on average 78.0\,arcsec$^2$ for the griW selection and 100.1\,arcsec$^2$ for the UgrizW selection.
The angular number density of the UgrizW selection also depends strongly on both SDSS and SCUSS seeing. This behaviour results from a correlation between the $U-r$ color term that enters the definition of the Fisher discriminant and those seeing parameters.

We model simultaneously the effect of every observational parameters using a multivariate regression which is quadratic as a function of the stellar density, the extinction and SDSS sky flux that exhibit explicit non-linear behaviour, and linear as a function of all other observational parameters. This modelling allows us to compute the predicted density of the selections in the absence of shot noise and cosmic variance given the value of the observational parameters.

The predicted density of the griW selection is contained within the $\pm7.5\%$ variation ($15\%$ total) as a function of the observational parameters, but the UgrizW selection is not, meaning that the UgrizW selection fails the homogeneity requirements of eBOSS. Based on our modelling, we propose a weighting procedure to remove the effect of the observational parameters on the number densities of both selections. 

We study the variation in number density of our two selections as a function of variations in imaging zero-points, showing that both selections are within the requirements of eBOSS.

We compute the large scale angular clustering of the two selections, showing that they both have excess of signal at, and above the BAO scale indicating the presence of systematic errors in the measurements. We calculate the cross-correlation between our selections and the observational parameters. For both selections, the cross-correlation with the stellar density has the largest signal followed by the cross-correlation with the extinction. For the UgrizW selection there are important cross-correlations with SDSS and SCUSS seeing, has expected from the previous results. 

Our weighting procedure successfully removes the signal in the cross-correlations between the selections and the observational parameters which represents an improvement of a factor of 30 in the case of the stellar density.

We compute and model the small scale angular clustering of the two catalogs in order to estimate the bias of the two selections. We find a bias of $1.35/D_a(z)$ for the griW selection and $1.28/D_a(z)$ for the UgrizW one.

Both catalogs are publicly available at https://data.sdss.org/sas/dr13/eboss/target/elg/fisher-selection/. 

\section*{Acknowledgements}

TD and JPK acknowledge support from the ERC advanced grant LIDA.
JC acknowledges financial support from MINECO (Spain) under project number AYA2012-31101.
AR acknowledges funding from the P2IO LabEx (ANR-10-LABX-0038) in the framework "Investissements d\textquoteright Avenir" (ANR-11-IDEX-0003-01) managed by the French National Research Agency (ANR).
JC and FP acknowledge support from the Spanish MICINNs Consolider-Ingenio 2010 Programme under grant MultiDark CSD2009-00064, MINECO Centro de Excelencia Severo Ochoa Programme under grant SEV- 2012-0249, and MINECO grant AYA2014-60641-C2-1-P.
EJ acknowledges financial support from CNRS/INSU, France.

Funding for SDSS-III has been provided by the Alfred P. Sloan Foundation, the Participating Institutions, the National Science Foundation, and the U.S. Department of Energy Office of Science. The SDSS-III web site is http://www.sdss3.org/.

SDSS-III is managed by the Astrophysical Research Consortium for the Participating Institutions of the SDSS-III Collaboration including the University of Arizona, the Brazilian Participation Group, Brookhaven National Laboratory, Carnegie Mellon University, University of Florida, the French Participation Group, the German Participation Group, Harvard University, the Instituto de Astrofisica de Canarias, the Michigan State/Notre Dame/JINA Participation Group, Johns Hopkins University, Lawrence Berkeley National Laboratory, Max Planck Institute for Astrophysics, Max Planck Institute for Extraterrestrial Physics, New Mexico State University, New York University, Ohio State University, Pennsylvania State University, University of Portsmouth, Princeton University, the Spanish Participation Group, University of Tokyo, University of Utah, Vanderbilt University, University of Virginia, University of Washington, and Yale University. 

Funding for the Sloan Digital Sky Survey IV has been provided by
the Alfred P. Sloan Foundation, the U.S. Department of Energy Office of
Science, and the Participating Institutions. SDSS-IV acknowledges
support and resources from the Center for High-Performance Computing at
the University of Utah. The SDSS web site is www.sdss.org.

SDSS-IV is managed by the Astrophysical Research Consortium for the 
Participating Institutions of the SDSS Collaboration including the 
Brazilian Participation Group, the Carnegie Institution for Science, 
Carnegie Mellon University, the Chilean Participation Group, the French Participation Group, Harvard-Smithsonian Center for Astrophysics, 
Instituto de Astrof\'isica de Canarias, The Johns Hopkins University, 
Kavli Institute for the Physics and Mathematics of the Universe (IPMU) / 
University of Tokyo, Lawrence Berkeley National Laboratory, 
Leibniz Institut f\"ur Astrophysik Potsdam (AIP),  
Max-Planck-Institut f\"ur Astronomie (MPIA Heidelberg), 
Max-Planck-Institut f\"ur Astrophysik (MPA Garching), 
Max-Planck-Institut f\"ur Extraterrestrische Physik (MPE), 
National Astronomical Observatory of China, New Mexico State University, 
New York University, University of Notre Dame, 
Observat\'ario Nacional / MCTI, The Ohio State University, 
Pennsylvania State University, Shanghai Astronomical Observatory, 
United Kingdom Participation Group,
Universidad Nacional Aut\'onoma de M\'exico, University of Arizona, 
University of Colorado Boulder, University of Portsmouth, 
University of Utah, University of Virginia, University of Washington, University of Wisconsin, 
Vanderbilt University, and Yale University.

The SCUSS is funded by the Main Direction Program of Knowledge Innovation of Chinese Academy of Sciences (No. KJCX2-EW-T06). It is also an international cooperative project between National Astronomical Observatories, Chinese Academy of Sciences and Steward Observatory, University of Arizona, USA. Technical supports and observational assistances of the Bok telescope are provided by Steward Observatory. The project is managed by the National Astronomical Observatory of China and Shanghai Astronomical Observatory.

This publication makes use of data products from the Wide-field Infrared Survey Explorer, which is a joint project of the University of California, Los Angeles, and the Jet Propulsion Laboratory/California Institute of Technology, and NEOWISE, which is a project of the Jet Propulsion Laboratory/California Institute of Technology. WISE and NEOWISE are funded by the National Aeronautics and Space Administration.




\bibliographystyle{mnras}
\bibliography{elg_technical} 








\bsp	
\label{lastpage}
\end{document}